\documentclass[11pt, oneside, a4paper]{article}

\usepackage{graphicx}
\usepackage{float}  
\usepackage{caption}
\usepackage{subcaption}

\usepackage{amsmath}
\usepackage{amsmath, esint}  
\usepackage{scalerel}  

\usepackage{hyperref}
\usepackage{color}
\usepackage{mathtools}
\usepackage{fixmath}
\usepackage{amssymb}
\usepackage{indentfirst}
\usepackage{setspace}
\usepackage{geometry}

\usepackage{authblk}
\usepackage{cite}

\hypersetup{
	colorlinks = true,
	urlcolor   = black,
	linkcolor  = black,
	citecolor  = black,
}

\title{A novel volume of fluid ghost-cell immersed boundary method for free surface flow interacting with structures}

\author[a,b]{Fan Chen}
\author[b]{Jinghua Wang}
\author[b,\thanks{\href{mailto:hf.duan@polyu.edu.hk}{Corresponding Author: H.F. Duan}}]{Huan-Feng Duan}

\affil[a]{School of Naval Architecture, Ocean and Civil Engineering, Shanghai Jiao Tong University,\newline Shanghai, 200240, China}
\affil[b]{Department of Civil and Environmental Engineering, The Hong Kong Polytechnic University,\newline Hong Kong SAR 999077, China}

\date{}

\begin{document}

\maketitle

\begin{abstract}
This paper presents a novel volume of fluid ghost-cell immersed boundary (IB) method for two-phase free surface flow interacting with structures. To circumvent the disturbance occurring around the intersection area of the IB and free surface when using the interpolation method for variable reconstruction, the fluid-structure interaction is firstly considered with the orthogonal IB by mimicking the imposition of boundary conditions in the body-conformal grid method. Treatments are subsequently performed to account for the non-orthogonal effect in accurately simulating the FSI, including the newly proposed flux-scaling and IB velocity re-evaluation methods. Further, a variable smoothing process and a flux correction method are adapted to handle moving boundary cases. Based on OpenFOAM, a two-phase flow solver has been developed. Both stationary and moving immersed boundary cases are used for validations. The numerical results reasonably agree with the corresponding laboratory data and other numerical simulation results, demonstrating the disturbance being effectively depressed and the solver’s accuracy in capturing fluid-structure interactions involving free surface flow.
\end{abstract}

\section{Introduction} \label{sec: introduction}
The Immersed Boundary Method (IBM) has gained much attention in the community of computational fluid dynamics (CFD) due to its advantages in simulating flows over complex geometries and being flexible in handling large-amplitude motions of structures \cite{Mittal_2005, SotiropoulosYang_2014, Zhang_2020}. Initially proposed by Peskin \cite{Peskin_1972}, it has been widely applied and developed over the past few decades. A large number of IBM variants have been proposed to deal with the Fluid-Structure Interaction (FSI) problems in many fields in both academia and industries \cite{Wang_2009, Constant_2017, KimChoi_2019, Tran_2020, Verzicco_2022, Chéron_2023}. However, relatively few studies have focused on IBM's application to problems involving two-phase flow interacting with structures \cite{LiuDing_2015}, which is particularly important in the context of free surface flow in coastal and ocean engineering \cite{ShenChan_2008, Luo_2024}.

To develop IBM for dealing with FSI problems involving free surface flow, suitable option(s) should be identified among these various types of IBMs. Generally, IBMs can be categorised into two groups, namely the continuous and discrete forcing IBMs, depending on how they handle the forcing term to simulate the FSI in the governing equations \cite{Mittal_2005, Jasak_2014}. The continuous forcing IBM considers analytical or ad-hoc forcing source terms before discretising the governing equations \cite{Mittal_2023}. Despite its simplicity, issues regarding the mass conservation of the fluid have been identified due to the inherently diffused immersed boundary (IB) \cite{Pan_2018}. Therefore, the continuous forcing IBM has limitations in the applications of general engineering problems. The discrete forcing IBM considers the FSI after discretising the governing equations. Sharp representation of the structure boundary can be considered and thus has the advantage of preserving mass conservation \cite{Mittal_2023, Pan_2018}. Depending on how the geometry elements are treated, it can be further classified into cut-cell IBM and ghost-cell IBM (GCIBM) \cite{Mittal_2023}. The cut-cell IBM is analogous to the Body-Conformal Grid Method (BCGM). The discretisation of governing equations relies on the fitted shape of the structure. It requires an efficient computational geometry module to obtain cut-cell information. However, implementing this method in three-dimensional (3D) cases can be challenging, especially for complex geometries \cite{Balaras_2004}. Moreover, some cut cells with small dimensions may be formulated, limiting the sizes of computational time steps. The GCIBM, in contrast, imposes boundary conditions (BCs) without requiring the cut cells’ geometry information \cite{Mittal_2008}. The FSI is achieved via polynomial interpolation approaches to reconstruct the forcing term or directly evaluate variables for BC imposition \cite{Jasak_2014, Mittal_2023, LiuHu_2014}. Some pioneering work (e.g., Mohd-Yusof \cite{Mohd-Yusof_1997} and Fadlun et al. \cite{Fadlun_2000}) has successfully applied GCIBM to simulate the two-phase flow \cite{LiuDing_2015, TsengFerziger_2003, OBrienBussmann_2018}. Therefore, this GCIBM becomes a good option to handle the free surface flow. 

In order to account for two-phase free surface flow interacting with structures, an IBM should be coupled with a two-phase flow modelling and integrated within a specific algorithm for solving governing equations. For modelling two-phase flow, the commonly used approaches include the volume of fluid (VOF), level-set (LS) and front tracking (FT). A number of studies have demonstrated the effectiveness of coupling these methods with IBMs \cite{Deen_2009, Zhang_2010, Calderer_2014}. Among these, FT is the Lagrangian method, requiring the updating of the markers of the interface. While this method is accurate in capturing the interface, its implementation can be challenging due to the potential complexity of interface evolution. Between these Eulerian approaches, VOF and LS, VOF is a scalar transport-based method without the requirement of recalculating the signed distance field when the interface changes, making it easier to implement and integrate with the Cartesian grid. 

Regarding the integration of IBM with the solving algorithms for governing equations, the fractional step method is commonly used to decouple the velocity-pressure. A number of studies have demonstrated the coupling of GCIBM with the fractional step method, where the forcing term reconstruction method is usually applied \cite{YangStern_2009, Zhang_2014, Lin_2016}. However, obtaining the forcing term may involve many terms and a series of mathematical derivations are required as demonstrated in \cite{YangStern_2009, Lin_2016}. In comparison, the GCIBM with variables directly evaluated is straightforward. However, few attempts have been made in this way, especially dealing with the free surface flow interacting with structures. Jasak et al. \cite{Jasak_2014} developed such GCIBM solvers within Foam-extend (one of the main forks of the widely used open-source CFD code OpenFOAM). The single-phase flow solver works reasonably well, but the proposed two-phase GCIBM lacks support in the modelling of two-phase flow interacting with moving IBs. Although some efforts have been made to address this, such as \cite{Chen_2020}, these solvers still face the disturbance issue around the intersection area of IB cells and the interface between the two phases \cite{Chen_2020, Jasak_2018}. The disturbance issue is shown as the smeared phase fractions and diffused velocity across the interface, leading to some instability around this area \cite{Jasak_2018}. It results in the mass conservation issue and affects the accuracy of FSI simulations. Upon investigating the implementation details of \cite{Jasak_2014} and referencing the related studies in GCIBM, the main reasons for the disturbance issue are considered as below:

i). Reconstructing variables using the polynomial fitting across the interface of a two-phase flow with a large density ratio may smooth the large gradients and lead to diffused variables near the IB \cite{Jasak_2018};

ii). Variables were reconstructed independently, neglecting the relation among them. This may lead to accumulated errors, e.g., if reconstructing a large pressure gradient at the IB based on the gradient field rather than directly calculating with the reconstructed pressure, the variable gets smoothed twice. 

To this point, the key challenge is the difficulty of finding an appropriate fitting method coupled with GCIBM to treat a two-phase flow with a large density ratio, while preventing the diffusion of variables around the IB. To overcome this challenge, this work proposes a novel GCIBM with the following innovative strategies: the two-phase FSI is firstly considered based on a “stair-step” IB by mimicking the FSI simulation in BCGM and then accounts for the effect of the actual structure boundary. The disturbance issue could thus be circumvented. Efforts are now required to consider the effect of the structure geometry details on the surrounding fluid. Additionally, methods such as handling dynamic mesh in BCGM could also be adapted to the GCIBM to deal with moving IBs. Besides, another advantage of the newly proposed model is its simple and straightforward imposition of BCs, compared with the existing two-phase GCIBMs based on forcing term reconstruction. 

With those in mind, this paper is organised as follows. The numerical method detailing the governing equations, discretised process, solving algorithm and IB treatment is given in Section \ref{sec: methodology}. The newly developed solver is verified first and then validated with both stationary and moving IB cases in Section \ref{sec: results}. This is followed by the most important conclusions summarised in Section \ref{sec: conclusions}.

\section{Numerical Method} \label{sec: methodology}
The proposed GCIBM is implemented within OpenFOAM (version 7.0). The solver development is based on its provided two-phase BCGM solver interFoam. The algorithm for solving governing equations is PIMPLE, which is a partitioned algorithm that combines the PISO (Pressure Implicit with Splitting of Operator) and SIMPLE (Semi-IMplicit Pressure-Linked Equation) algorithms \cite{Gatin_2017}. This algorithm is suitable for transient flows with relatively large Courant numbers and can achieve a good balance between accuracy and computational efficiency \cite{Loginov_2014}. 

The two-phase free surface flow (i.e., the air and water in this work) is assumed to be transient, incompressible, immiscible and isothermal. The governing equations and their discretisation using the finite volume method (FVM), along with PIMPLE without the IB consideration are first given in Section \ref{sec: eqs}. The IB recognition and variable reconstruction are presented in Section \ref{sec: IB}. The coupling of GCIBM and PIMPLE is detailed in Section \ref{sec: GCIBM}, where several novel treatments for capturing FSI are implemented to avoid the disturbance issue. The solving process is briefly summarised in Section \ref{sec: summary}.

\subsection{Governing equations discretisation} \label{sec: eqs}
The governing equations of the free surface flow are Navier-Stokes equations (NSEs):
\begin{equation} \label{eqn: continuity}
	\nabla\cdot{\bf u} = 0,	
\end{equation}
\begin{equation} \label{eqn: orgMomentum}
	{\partial{\rho\bf u}\over{\partial t}} + \nabla\left(\rho{\bf u}{\bf u}\right)= -\nabla p + \nabla\left[\mu\left(\nabla{\bf u}\right)\right] + {\bf g} +{\bf f}_s,
\end{equation}

\noindent where the variables to be resolved are the fluid velocity vector {\bf u}  and pressure scalar {\it p}. The variable $\rho$ indicates the fluid density, $\mu$ the fluid dynamic viscosity, {\bf g}  the gravitational vector and {\bf f}$_s$ denotes the surface tension. The air-water interface is updated by solving a transport equation about phase fraction $\alpha$ \cite{HirtNichols_1981}: 
\begin{equation} \label{eqn: orgAlphaEqn}
	{\partial{\alpha}\over{\partial t}} + \nabla\cdot\left({\bf u}\alpha\right)= 0.
\end{equation}

\noindent In order to integrate the VOF method with NSEs, the Continuum Surface Force model \cite{Francois_2006, Baltussen_2014} is introduced to obtain {\bf f}$_s$:
\begin{equation} \label{eqn: fsEqn}
	{\bf f}_s = \sigma_{s}\kappa\alpha.
\end{equation}

\noindent In Eq. (\ref{eqn: fsEqn}), $\sigma_s=$ 0.07 N/m is the surface tension coefficient of air and water and $\kappa = -\nabla\cdot{\bf n}$ is the curvature of the air-water interface with {\bf n}  denoting the interface normal vector. By dividing p into dynamic $p_d$ and static components $\rho{\bf g}\cdot{\bf h}$ and taking them into Eq. (\ref{eqn: orgMomentum}), the momentum equation becomes \cite{Jasak_1996}
\begin{equation} \label{eqn: momentum}
	{\partial{\rho\bf u}\over{\partial t}} + \nabla(\rho{\bf u}{\bf u}) - \nabla\cdot(\nu\nabla{\bf u}) - \nabla{\bf u}\cdot\nabla\nu= 
	-\nabla p_d + {\bf g}\cdot{\bf h}\nabla\rho + \sigma_{s}\kappa\alpha,
\end{equation}

\noindent Equations (\ref{eqn: continuity}), (\ref{eqn: orgAlphaEqn}) and (\ref{eqn: momentum}) are the governing equations for the free surface flow.

Within the PIMPLE framework, Eq. (\ref{eqn: orgAlphaEqn}) responsible for the update of the air-water interface is discretised via FVM as:
\begin{equation} \label{eqn: alphaEqn}
	{{\rm\Delta} V}{{\alpha^{\rm n+1}-\alpha^{\rm n}}\over{{\rm\Delta} t}} + \sum{F{\alpha_f}^{\rm n+1}} = 0,
\end{equation}

\noindent where the superscript n denotes a time instant, subscript {\it f} the face variable, ${\rm\Delta} t$ the time interval and ${\rm\Delta} V$ indicates the cell volume. The face flux F is the product of a face’s velocity vector {\bf u}$_f$ and its area vector {\bf S}$_f$. The implicitly resolved $\alpha$ is then corrected explicitly with high-order accuracy fluxes by the MULES (Multidimensional Universal Limiter for Explicit Solution) limiter, which is commonly used to retain a scalar’s boundness based on the flux correction method \cite{McGinn_2022}. Variables such as $\rho$ and $\mu$ are then evaluated with the updated $\alpha$ prior to solving Eqs. (\ref{eqn: continuity}) and (\ref{eqn: momentum}). In OpenFOAM, the discrete form of Eqs. (\ref{eqn: continuity}) and (\ref{eqn: momentum}) are decoupled into the velocity equation and the Pressure Poisson Equation (PPE) as below \cite{Jasak_1996, OpenFOAMv7}: 
\begin{equation} \label{eqn: UEqn}
	A_P{\bf u} = S_P - \sum{A_N{\bf u}_N} = H,
\end{equation}
\begin{equation} \label{eqn: PPE}
	\nabla\cdot\nabla\left(r_{A(u)}{p_d}\right) = \nabla\cdot{\bf u}.
\end{equation}

\noindent The term $r_{A(u)}$ is the reversed of diagonal coefficient $A_P$ in the matrix corresponding to discretised velocity equation. The coefficient $A_P$ contains the cell’s own effect on the flow field and the effect from its neighbour cell N is denoted with $A_N$. The term H involves the source term and other variables obtained explicitly from the previous time step. $A_P$ is crucial for the subsequent solving process. Equations (\ref{eqn: UEqn}) and (\ref{eqn: PPE}) can be solved based on the PISO loop in the PIMPLE algorithm \cite{OpenFOAMv7, Gimenez_2019}:
\begin{equation} \label{eqn: PIMPLE_UEqn}
	{\bf u}^* = {H^*\over{A_P^*}} + {{\bf g}\cdot{\bf h}\nabla\rho^*+{\bf f}_s-\nabla{p_d^*}\over{A_P^*}},
\end{equation}
\begin{equation} \label{eqn: PIMPLE_PPE}
	\nabla\cdot\nabla\left(r_{A(u)}^*{p_d^{**}}\right) = \nabla\cdot{\bf u}^*,
\end{equation}
\begin{equation} \label{eqn: PIMPLE_UFinal}
	{\bf u}^{**} = {\bf u}^* - r_{A(u)}^*\nabla{p_d^{**}}.
\end{equation}

\noindent Equations (\ref{eqn: PIMPLE_UEqn}-\ref{eqn: PIMPLE_UFinal}) can be solved repeatedly, where the superscript * and ** denote the intermediate time steps between the n and n+1  time instants. In this work, since the temporal term is discretised using the implicit Euler scheme and the Gauss linear scheme is employed in the discretisation of convection and dispersion terms in Eq. (\ref{eqn: momentum}), $A_P^*$ and $H^*$ are thus given as:
\begin{equation} \label{eqn: calcAp}
	A_P^* = {1\over r_{A(u)}^*} = {\rho^*\over{\rm\Delta} t} + {\rho^*\over{\rm\Delta} V}\sum{F^*} + {\rho^*\over{\rm\Delta} V}\sum\nu^*{\left|{\bf S}_f\right|\over d},
\end{equation}
\begin{equation} \label{eqn: calcH}
	H^* = {\rho^*}{{\rm\Delta} V}{{\bf u}^*\over{{\rm\Delta} t}} - \sum{A_N^*{\bf u}_N^*}.
\end{equation}

\subsection{IB recognition and variable reconstruction} \label{sec: IB}
\begin{figure}[H]
	\centering
	\includegraphics[]{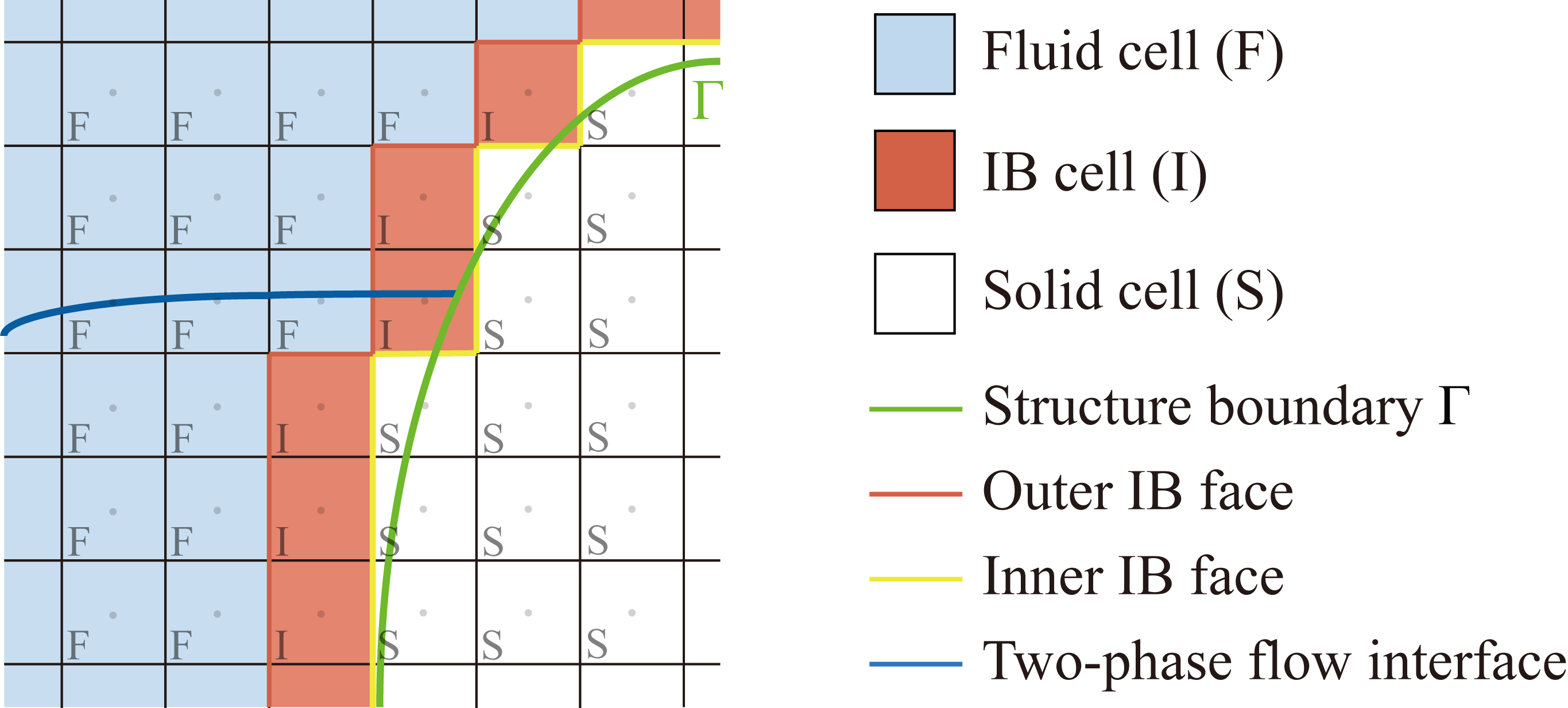}  
	\caption{Illustration of the cell identity and IB-related concepts}
	\label{fig:IB}
\end{figure}

The structure is represented with the STL (STereoLithography) format, where the structure surface is described with a series of triangular faces. The triangular faces’ vertices serve as Lagrangian markers of a structure boundary in the computational domain. Based on the relationship of a cell’s centre location with respect to the structure boundary, cells are classified into solid, IB and fluid cells, as illustrated in Fig. \ref{fig:IB}. The cells whose geometry centres are located within the structure are solid cells and they are supposed to have no influence over the fluid. The effect of the structure on the surrounding fluid is achieved by taking the special treatment via the IB cells, defined as cells one layer outside these solid cells. The remaining cells belong to fluid cells. To facilitate the evaluation of variables such as flux, some faces of IB cells are further classified into two groups: the inner and outer IB faces, based on their grid topologies. The inner IB faces connect IB cells and solid cells, while the outer IB faces are located between IB cells and fluid cells. In order to achieve the above cell or face classification, A volume-based approach provided by OpenFOAM is employed to determine whether a cell centre is inside or outside the structure surface \cite{OpenFOAMv7}.

In this work, Dirichlet (for given values) and Neumann (for given gradients) BCs are considered, which have already covered a wide range of FSI problems. The variables of IB cells are reconstructed via the moving least squares (MLS) approach, as demonstrated in \cite{LiuHu_2014, VanellaBalaras_2009}. For an IB cell variable $\phi_{\scaleto{IB}{4.5pt}}$, it is evaluated with
\begin{equation} \label{eqn: MLS}
	\phi_{\scaleto{IB}{4.5pt}} = c_0 + \sum\limits_{\rm k=1}^{{\rm N}_c}{c_{\rm k} p_{\rm k}}
\end{equation}
where $c_0$ and $c_{\rm k}$ are the interpolation coefficients, $p_{\rm k}$ the polynomial function and ${\rm N}_c$ indicates the number of $p_{\rm k}$. For Dirichlet BC, $c_0$ can be specified directly with the given value $\phi_{\Gamma}$ with the subscript $\Gamma$ denoting the variables at the structure boundary. For Neumann BC, $c_0$ is to be resolved with the given gradient ($\nabla\phi_{\Gamma}$ and in this case $\phi_{\scaleto{IB}{4.5pt}}$ should satisfy 
\begin{equation} \label{eqn: MLS_Neumann}
	{\partial\phi_{\scaleto{IB}{4.5pt}}\over\partial{\bf n}_{\scaleto{IB}{4.5pt}}} = \nabla\left(\sum\limits_{\rm k=1}^{{\rm N}_c}{c_{\rm k} p_{\rm k}}\right)\cdot{\bf n}_{\scaleto{IB}{4.5pt}} = \left(\nabla\phi\right)_\Gamma,
\end{equation}

\noindent where {\bf n}$_{\scaleto{IB}{4.5pt}}$ is a unit vector pointing from the IB cell centre to its nearest point (defined as IB point) at the structure boundary. Given the reconstruction with MLS approach has been presented in a number of studies, details of how to obtain $c_{\rm k}$ are presented in Appendix A.

Note that when reconstructing the bounded scalar $\alpha$, a restriction function should be applied to ensure its boundness, as well as to help retain the sharpness of the air-water interface. Some studies have demonstrated the effectiveness of applying the smoothed Heaviside function (e.g. in \cite{OBrienBussmann_2018, YangStern_2009}) and Sigmoid function \cite{Kim_2021} to confine the diffusion of a scalar to a certain width. Here, a Sigmoid function is applied
\begin{equation} \label{eqn: restrictAlpha}
	\alpha_{IB} = {1\over{1 + e^{-\varepsilon\left(\alpha^{MLS}-0.5\right)}}},
\end{equation}

\noindent where $\varepsilon$ is a constant and $\alpha^{MLS}$ is the reconstructed value via MLS.

\subsection{Coupled GCIBM and PIMPLE algorithm for free surface flow} \label{sec: GCIBM}
This section details the coupling of GCIBM with PIMPLE to simulate the interaction between the free surface flow and the IB. To prevent the aforementioned disturbance, the BC is initially imposed on orthogonal IB (structure surface represented with “stair step” boundary) by mimicking the FSI simulation in BCGM. When the moving IB is involved, some dynamic mesh handling methods in BCGM are adapted to help account for the moving IB case. Since the IB has been treated orthogonally thus far, treatments are subsequently performed in consideration of the non-orthogonal effect to accurately simulate the FSI. The following subsections are organised in corresponding to each module’s appearance order in the solving processing.
\subsubsection{BC imposition on orthogonal IB} \label{sec: orthogIB}
The orthogonal IB is the boundary consisting of inner IB faces as shown in Fig. \ref{fig:IB}. It is initially regarded as the structure boundary in the solving process. The BC imposition mimics that in BGCM, involving the evaluations of boundary values and the boundary faces’ corresponding matrix coefficients. Therefore, according to the given BCs, the inner IB face values can be directly evaluated. It should be noted that a number of terms need to be calculated prior to the construction of the governing equations, however, there is no IB consideration in the calculation of terms when using the originally provided functions in OpenFOAM, i.e., the gradients are obtained by treating all internal faces as fluid faces instead of IB faces. These gradients should be corrected to consider the presence of the IB. According to the PIMPLE solving algorithm, the correction involves the gradients of $\alpha$, $\rho$, {\bf u}, and $p_d$. Since the gradient calculation is based on the Gauss Theorem \cite{OpenFOAMv7}, the corrected value at an IB cell (denoted with subscript IB) is thus given as:
\begin{equation} \label{eqn: corrGrad}
	(\nabla\phi)_{\scaleto{IB}{4.5pt}} = {\oiint\phi d{\bf S} \over{V_{\scaleto{IB}{4.5pt}}}} = {1\over{V_{\scaleto{IB}{4.5pt}}}}\left(\sum_{{\rm N}_{fluid}}\phi_f{\bf S}_f + \sum_{{\rm N}_{\scaleto{IB}{4.5pt}}}\phi_{f, inner}{\bf S}_{f, inner}\right),
\end{equation}

\noindent where the subscripts N$_fluid$ and N$_IB$ represent the numbers of the fluid and inner IB faces within an IB cell, respectively. 

As the inner IB faces are now treated as the structure boundary, these faces’ corresponding matrix coefficients need to be corrected accordingly, by cutting off the connection of IB cells with their neighboured fluid cells. Therefore, the diagonal coefficient containing inner IB faces is corrected with 
\begin{equation} \label{eqn: corrAp}
	A_{P, IB}=A_P-\sum_{{\rm N}_{\scaleto{IB}{4.5pt}}}\chi_{\scaleto{IB}{4.5pt}}=\sum_{{\rm N}_{fluid}}\chi_{fluid},
\end{equation}

\noindent where $A_P$ is the calculated value based on Eq. (\ref{eqn: UEqn}) without the consideration of the IB, $\chi$ denotes the contribution coefficient of a (fluid or inner IB) face in the matrix formed by the discretised equation. The source term should also exclude the component associated with the inner IB face’s variable \cite{Pan_2018}, which will be presented in the following subsections.
\subsubsection{Variable smoothed for identity changed cells} \label{sec: smooth}
The correction method presented in Section \ref{sec: orthogIB} can be directly applied in the stationary IB case to simulate FSI. However, additional treatments should be employed in the moving IB case, as a cell’s identity at the current time step may be different from that at the previous one. There will be some unphysical values used in the evaluation if using MLS directly. For instance, an IB cell in the last time step now becomes a fluid cell, $\phi_{\scaleto{IB}{4.5pt}}$ is involved in the MLS reconstruction rather than the expected $\phi_{fluid}$. As a result, it is necessary to consider the changed variable due to the cell identity transition. A variable smoothing process is performed, similar to the topological changes (re-meshing) method in BCGM. Given that new and old IB cells should be re-evaluated, two-layer cells are to be smoothed using the MLS approach, i.e., the current IB cells and one layer immediately outside these cells. Since the values of the last time step at the IB are known, MLS is only applied with Dirichlet type. 

According to the solving process shown in Section \ref{sec: eqs}, {\bf u} and $\alpha$ are required to be smoothed at the beginning of each time step. The related coefficient $r_{A(u)}$  in the equation should also be recalculated accordingly. Since $r_{A(u)}$  may have a large gradient across the air-water interface, directly reconstructing $r_{A(u)}$ can smooth the gradient. Thus, $r_{A(u)}$ at an IB cell is obtained with the reconstructed {\bf u}, the $\alpha$-related $\rho$ and kinematic viscosity $\nu$ based on Eq. (\ref{eqn: calcAp}): 
\begin{equation} \label{eqn: calcRAU}
	r_{A(u), IB}^{\rm n} = 
	{\rho_{\scaleto{IB}{4.5pt}}\over{{\rm\Delta} V}}\left({{\rm\Delta} V\over{{\rm\Delta} t}} + \sum{\bf u}_f^{\scaleto{IB}{4.5pt}}\cdot{\bf S}_f + \sum\nu_f^{\scaleto{IB}{4.5pt}}{\left|{\bf S}_f\right|\over d}\right)^{-1},
\end{equation}

\noindent where the superscript IB denotes the variable involving the IB-related value.

\subsubsection{Flux correction for moving IB} \label{sec: corrPhi}
In BCGM, the moving structure boundary leads to boundary face/cell displacements and then the rest of the grid points move according to a certain algorithm to accommodate the boundary displacement. The changed locations of the grid points result in face fluxes in the convection term in the momentum equation \cite{Jasak_2009, Ferziger_2019}. Since the background mesh does not change in the IBM, the fluxes caused by the moving IB only occur at IB cells. The resulting fluxes among the remaining cells should be obtained, similarly, by solving the Poisson equation about a scalar $p^{corr}$ \cite{OpenFOAMv7}. This flux correction process is given below
\begin{equation} \label{eqn: pcorrEqn}
	\nabla\cdot\left(r_{A(u)}^{\scaleto{IB}{4.5pt}}\nabla p^{corr}\right) = \nabla\cdot{\bf u}^{\scaleto{IB}{4.5pt}},
\end{equation}
\begin{equation} \label{eqn: corrPhi}
	F^{corr} = F - p^{corr},
\end{equation}

\noindent where the term $\nabla\cdot{\bf u}^{\scaleto{IB}{4.5pt}}$ contains two components: $\nabla\cdot{\bf u}^{fluid}$ obtained by the previous time step at non-IB cells and the velocity divergence involving {\bf u}$_{f, inner}$ at IB cells. In Eq. (\ref{eqn: corrPhi}), the $F$ and $F^{corr}$ are the fluxes prior to and after correction, respectively. According to Eq. (\ref{eqn: calcRAU}), there is a significant difference in values of $r_{A(u)}$ for a two-phase flow with a large density ratio. The term $r_{A(u)}$ thus plays an important role in preventing the diffusion of variables (i.e., to avoid the smoothed value caused by MLS) around the intersection area of free surface and IB.

\begin{figure}[H]
	\centering
	\includegraphics[width=0.9\textwidth]{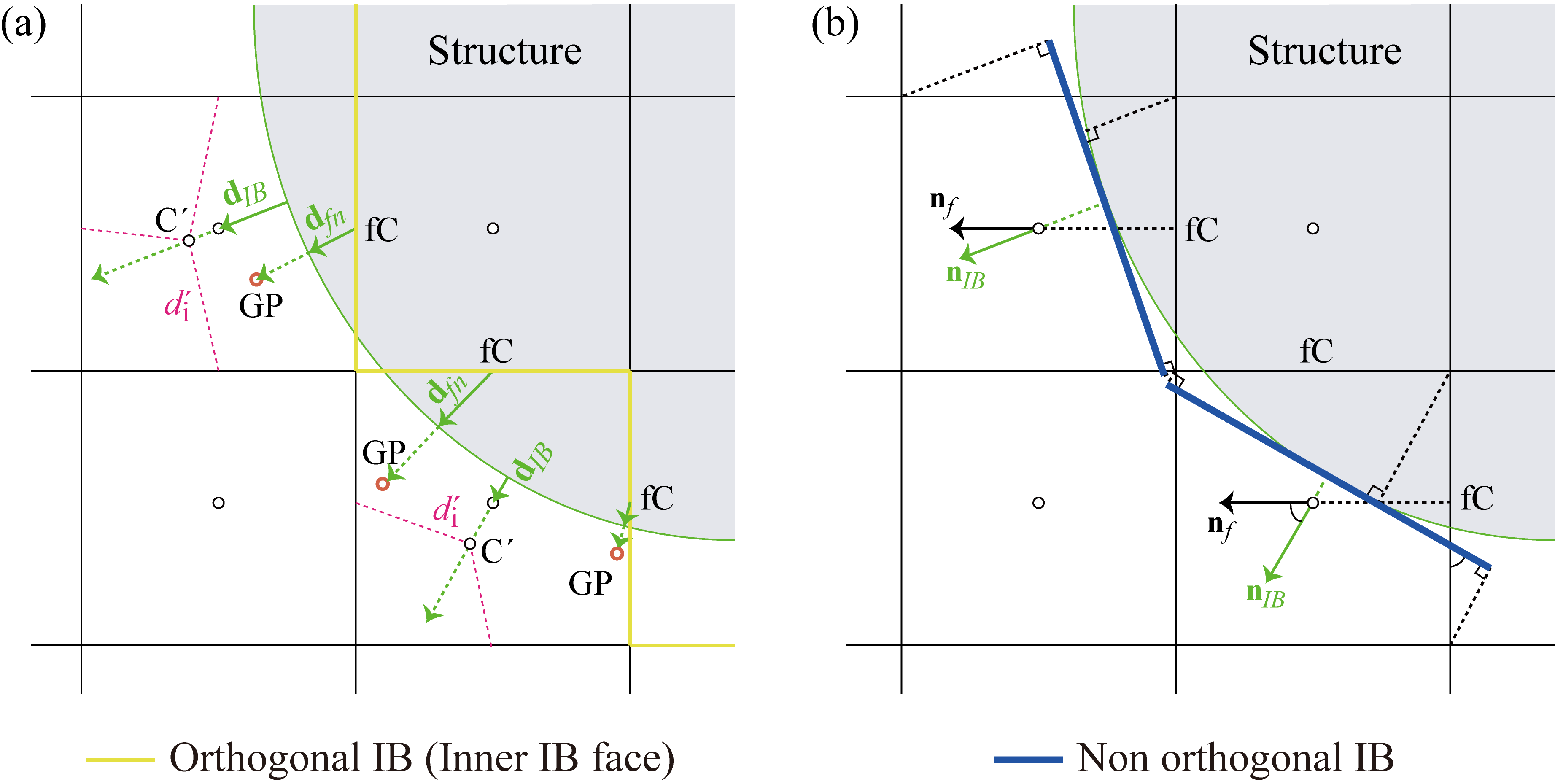}
	\caption{Schematic showing the (a) ghost points (GPs) and (b) non-orthogonal IB}
	\label{fig:IB_schematic}
\end{figure}

Considering the centres of the inner IB faces may be located within the structure boundary or an IB cell, a ghost point (GP) should be determined and used for the evaluation as illustrated with a two-dimensional (2D) case in Fig. \ref{fig:IB_schematic}. The position vector of this ghost point {\bf X}$_{GP}$ is used in the determination of $p_{\rm k}$ in Eq. (\ref{eqn: MLS}), similar to the image point treatment in other GCIBMs such as \cite{Chi_2020, Vanna_2020}
\begin{equation} \label{eqn: GPLocation}
	{\bf X}_{GP} = {\bf X}_{fC} + 2{\bf d}_n,
\end{equation}

\noindent where {\bf d}$_n$ is the vector pointing from the inner IB face centre (fC) to its nearest point on IB. To decrease the velocity difference between the reconstructed {\bf u}$_{f, inner}^{MLS}$ and the structure boundary velocity on the corresponding IB point {\bf u}$_\Gamma$, {\bf u}$_{f, inner}$ can be evaluated as the averaged value:
\begin{equation} \label{eqn: uf}
	{\bf u}_{f, inner} = {{{\bf u}_{f, inner}^{MLS} + {\bf u}_\Gamma}\over 2},
\end{equation}

\noindent Subsequently, the corrected $F^{corr}$ is used to update the air-water interface by solving Eq. (\ref{eqn: alphaEqn}), prior to PISO loop for velocity-pressure decoupling.

\subsubsection{Treatments for non-orthogonal IB} \label{sec: nonorthogIB}
Given the actual structure boundary may not be parallel to the cell faces, the effect of the non-orthogonal boundary on the fluid should be considered for accurately capturing FSIs. In this section, treatments to account for this non-orthogonal effect and the IB variable re-evaluation are presented. Similar to BCGM, the non-orthogonal effect is considered in solving PPE within PISO loop. With the IB related terms, the discrete pressure equation Eq. (\ref{eqn: PPE}) is given below 
\begin{equation} \label{eqn: discretePPE}
	\left(-\sum_{{\rm N} - {\rm N}_{\scaleto{IB}{4.5pt}}}r_{A(u), f}^* \right)p_d^{**} + \sum_{{\rm N} - {\rm N}_{\scaleto{IB}{4.5pt}}}r_{A(u), f}^*p_{d, N}^{**} = \nabla\cdot{\bf u}^{IB, *}.
\end{equation}

\noindent In Eq. (\ref{eqn: discretePPE}), r$_{A(u), f}^*$ is interpolated $A_P^{\scaleto{IB}{4.5pt}}$ from the cell to the face. Similar to Eq. (\ref{eqn: pcorrEqn}), $\nabla\cdot{\bf u}^{IB, *}$ contains the components $\nabla\cdot{\bf u}_{fluid}^*$ at fluid cells and $\nabla\cdot{\bf u}_{\scaleto{IB}{4.5pt}}^*$ at IB cells. The term $\nabla\cdot{\bf u}_{fluid}^*$ can be obtained from the corrected velocity equation Eq. (\ref{eqn: UEqn}), whose discretised form is
\begin{equation} \label{eqn: discreteUEqn}
	A_P^{\scaleto{IB}{4.5pt}}{\bf u}^* + \sum_{{\rm N} - {\rm N}_{\scaleto{IB}{4.5pt}}}A_{N, f}{\bf u}_N^* = S_P^{IB, *}.
\end{equation}

\noindent In Eq. (\ref{eqn: discreteUEqn}), $S_P^{IB, *}$ indicates the reconstructed IB variables have been involved in the source term. 

For the calculation of $\nabla\cdot{\bf u}_{\scaleto{IB}{4.5pt}}^*$, the consideration of non-orthogonal effect should be involved, as $\nabla\cdot{\bf u}_{\scaleto{IB}{4.5pt}}^*$ directly affects resolved boundary pressure. The structure boundary is approximated with a non-orthogonal IB here, which is the projection of the actual structure boundary onto the tangent plane of the vector {\bf n}$_{\scaleto{IB}{4.5pt}}$ as illustrated in Fig. \ref{fig:IB_schematic}(b). An IB cell’s corresponding non-orthogonal IB is a line in 2D or a plane in 3D cases. The inner IB face velocity {\bf u}$_{f, inner}^*$ is thus directly specified with {\bf u}$_\Gamma$. In order to account for the non-orthogonal effect, a flux scaling approach is proposed: the magnitude of the inner IB face’s area vector is scaled based on the face’s projection onto this non-orthogonal IB (Fig. \ref{fig:IB_schematic}b). With the scaled flux, the velocity divergence $\nabla\cdot{\bf u}^{IB, *}$ is thus given as
\begin{equation} \label{eqn: scaledFlux}
	\nabla\cdot{\bf u}^{IB, *} = \nabla\cdot{\bf u}_{fluid}^{*} + \sum_{{\rm N}_{\scaleto{IB}{4.5pt}}}{{\bf u}_\Gamma\cdot{\bf S}_f \over V_{\scaleto{IB}{4.5pt}}}\left|{\bf n}_{\scaleto{IB}{4.5pt}}\cdot{\bf n}_{f, inner}\right|.
\end{equation}

After solving Eq. (\ref{eqn: discretePPE}), a relaxation factor is applied to update $p_d^{**}$ with the value obtained from the previous iteration, to help suppress the pressure oscillation in moving IB cases. Subsequently, {\bf u}$^{**}$ at the fluid cells can be obtained using Eq. (\ref{eqn: PIMPLE_UFinal}). Similar to the evaluation of velocity using weighted interpolation at the end of PISO loop in BCGM solvers, the IB cell’s velocity {\bf u}$_{\scaleto{IB}{4.5pt}}^{**}$ is finally re-evaluated to account for the effect the non-orthogonal IB. Here, a commonly used inverse distance weighted method is applied \cite{Choung_2021}:
\begin{equation} \label{eqn: re_evaluateU}
	{\bf u}_{\scaleto{IB}{4.5pt}}^{**} = {
		{ {1\over{|{\bf d}_{\scaleto{IB}{4.5pt}}|}}{\bf u}_\Gamma + \sum_{{\rm i} = 1}^{{\rm N} - {\rm N}_{\scaleto{IB}{4.5pt}}}{1\over{d_i}}{\bf u}_{f, i}}
		\over
		{{1\over{|{\bf d}_{\scaleto{IB}{4.5pt}}|}} + \sum_{{\rm i} = 1}^{{\rm N} - {\rm N}_{\scaleto{IB}{4.5pt}}}{1\over{d_i}}}
	}.
\end{equation}

\noindent In Eq. (\ref{eqn: re_evaluateU}), the distance $d_i$ is calculated based on the IB cell’s new centre. Note that accurately obtaining its location requires a series of intersection determinations and may need some computational resources, especially when the geometry is complex in 3D cases. This new centre of the IB cell is approximated here. Given a special case when the non-orthogonal IB is parallel to a coordinate axis or plane, the new IB cell centre can be easily obtained. The same way is applied to calculate the approximated IB cell centre by assuming it is rotated from the special (orthogonal IB) case, as illustrated in Fig. \ref{fig:IB_schematic}(a):
\begin{equation} \label{eqn: newCellCentre}
	{\bf X}_C^\prime = {\bf X}_\Gamma + {1\over2}\left(1+{{\rm\Delta} l\over|{\bf d}_{\scaleto{IB}{4.5pt}}|}\right){\bf d}_{\scaleto{IB}{4.5pt}}.
\end{equation}
\noindent Here, {\bf d}$_{\scaleto{IB}{4.5pt}}$ is the vector from an IB point to the IB cell’s centre and ${\rm\Delta} l$ is the IB cell’s averaged size.

\subsection{Summary of the solving process} \label{sec: summary}
The proposed GCIBM has been developed as a new solver, {\it GCIBMFoam}, within OpenFOAM. In order to avoid the disturbance issue, the variable values and matrix coefficients are firstly corrected based on the orthogonal IB, by mimicking the FSI simulation in BCGM. Then based on the non-orthogonal IB, a newly proposed flux-scaling method is employed and the IB cell velocity is re-evaluated to accurately capture the FSI. To support the FSI simulation including moving structures, auxiliary modules have also been developed, which are presented in Appendix B.

\begin{figure}[H]
	\centering
	\includegraphics[width=\textwidth]{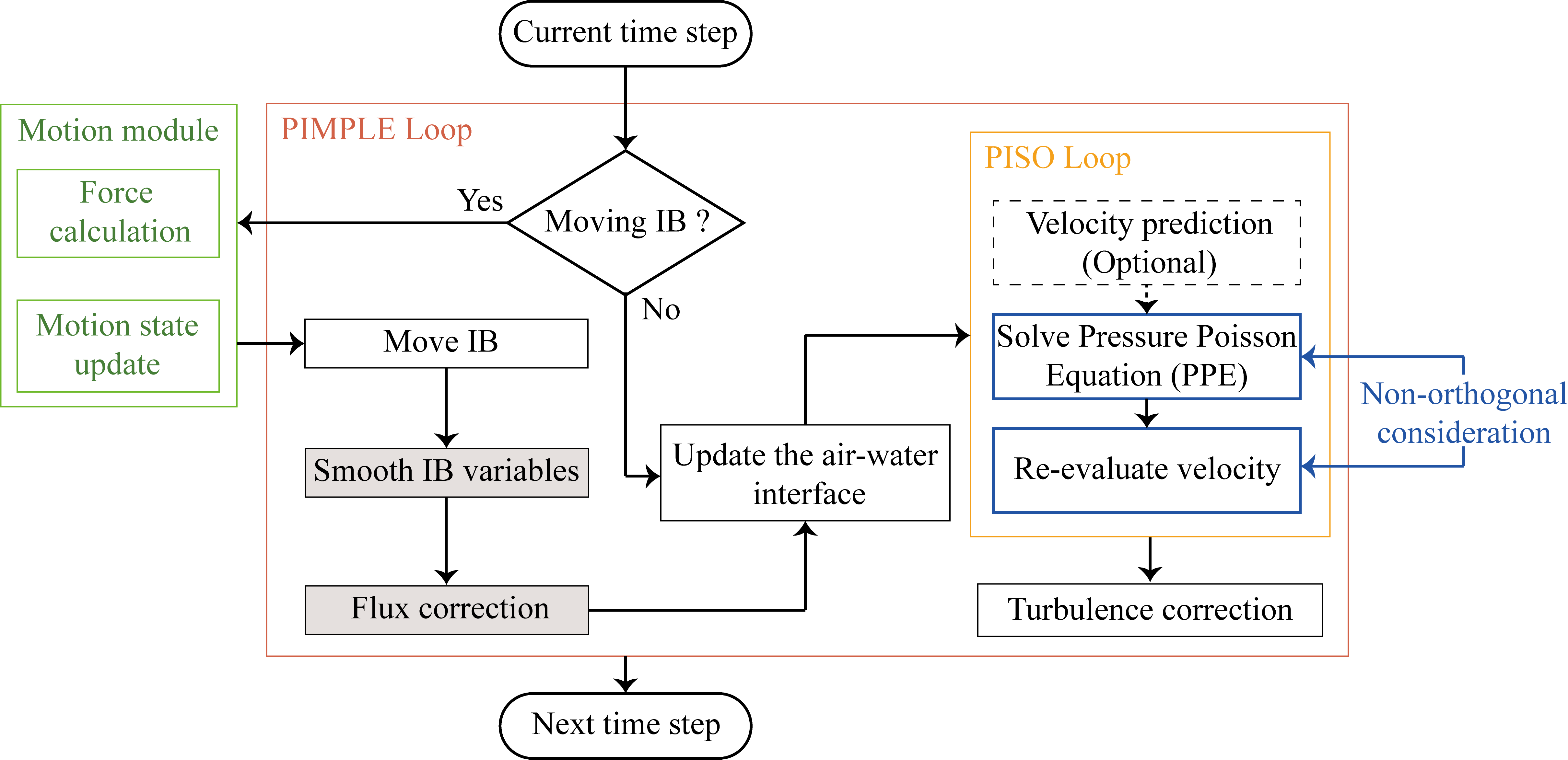}
	\caption{Flowchart of the GCIBM solver within one time step}
	\label{fig:flowchart}
\end{figure}

The flowchart illustrating the solving process within one time step is shown in Fig. \ref{fig:flowchart}. For the stationary IB case, the air-water interface is directly updated by solving Eq. (\ref{eqn: alphaEqn}). For the moving IB case, prior to solving the phase fraction equation, a variable smoothing process using the MLS approach, Eq. (\ref{eqn: MLS}), is performed to smooth the solutions to avoid unphysical variables due to changed cell identities. The fluxes are then corrected using Eqs. (\ref{eqn: pcorrEqn}) and (\ref{eqn: corrPhi}), which are analogous to the correction in the moving mesh method. After this, in both the stationary and moving IB cases, the velocity-pressure is decoupled within the PISO loop. The reconstructed cell and face values are used in the corrections of gradients and matrix coefficients. To consider the non-orthogonal effect, a flux-scaling method is proposed and applied with Eq. (\ref{eqn: scaledFlux}). After obtaining the velocity at the new time step, its re-evaluation is performed based on the non-orthogonal IB via Eq. (\ref{eqn: re_evaluateU}). All the corrections and reconstructions of IB-related variables and terms involved in the above solving process are summarised in Table \ref{table:variables}. 

At the end of each PIMPLE loop, the velocity can be corrected to model the turbulence. However, the current implementation does not involve reconstructing velocity within the boundary layer. A similar treatment of IB as \cite{Nguyen_2020} is employed by assuming the turbulent kinematic and its dissipation rate are zero at the inner IB faces.
\begin{table}[H]
	\centering
	\caption{List of variables or gradients to be reconstructed or corrected}
	\label{table:variables}
	\begin{tabular}{l|l}
		\hline
		IB variable reconstruction & $\alpha$ (only for moving IBs), {\bf u}, $p_d$ \\
		\hline
		IB-related value correction & $\alpha$, {\bf u}, $p_d$, $r_{A(u)}$ \\
		\hline
		IB-related gradient correction\quad\quad & $\nabla\alpha$, $\nabla{\bf u}$, $\nabla\rho$, $\nabla p_d$ \\
		\hline
		Equation correction & $\alpha$, {\bf u}, $p_d$, $p^{corr}$ \\ 
		\hline
				
	\end{tabular}
\end{table}

\section{Numerical Results} \label{sec: results}
This section presents the solver verification first, followed by the order of accuracy analysis. Then the solver is validated with canonical FSI problems, which are organised according to whether the IB is stationary or moving. The selected cases along with their purposes are listed below.

$\bullet$ Dam break flow over a semi-cylinder: verification of the newly proposed GCIBM in avoiding the disturbance issue and well preserving the water mass (Section \ref{sec: verification}).

$\bullet$ Water exit of a cylinder with prescribed motion: investigate the order of accuracy of the newly developed GCIBM solver (Section \ref{sec: accuracy})

$\bullet$ Solitary wave past a block: validate the solver with the case involving the wave-structure interaction (Section \ref{sec: fixedIB})

$\bullet$ Water entry of a freely falling wedge: validate the solver with the 2D case involving the resolved structure motion (Section \ref{sec: waterEntry})

$\bullet$ Tsunamis generated by iceberg calving: further validate the solver and examine the flow field evolution of 3D moving IB cases with the resolved structure motion (Section \ref{sec: IBT})

\subsection{Solver verification} \label{sec: verification}
\begin{figure}[H]
	\centering
	\includegraphics[width=\textwidth]{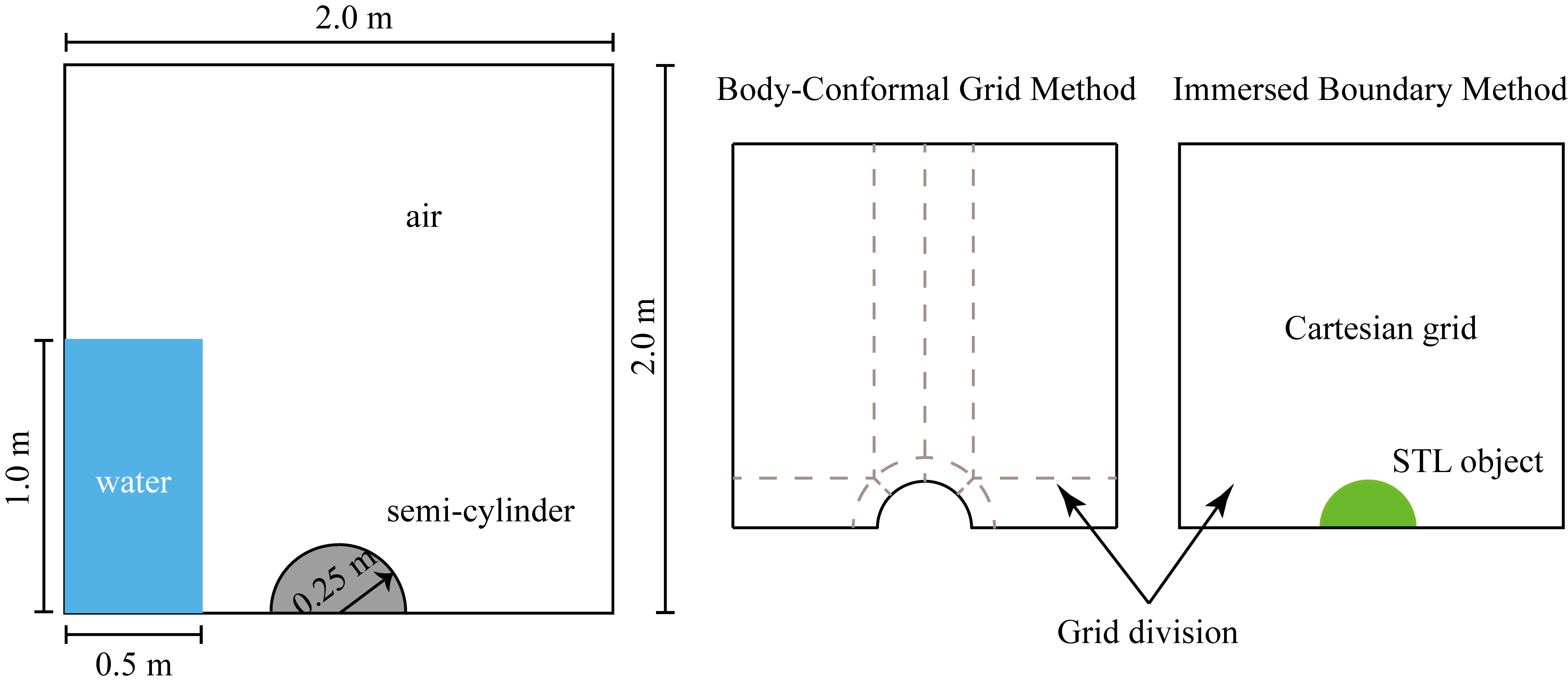}
	\caption{Sketch of the numerical setup of dam break flow over a semi-cylinder, together with the grid division in the body-fitted grid approach and IBM}
	\label{fig:setup_damBreak}
\end{figure}

Mass conservation is an important aspect of a two-phase flow solver's performance. In this section, the water volume in a dam break flow over a semi-cylinder case (2D) is examined and compared with that obtained by the BCGM solver interFoam and the IBM solver {\it interIbFoam} (provided in Foam-extend version 4.0). The flow field evolutions obtained by these three solvers are also presented to illustrate the effect of the disturbance and help to verify its circumvention by the newly developed solver. The numerical setup is illustrated in Fig. \ref{fig:setup_damBreak}. The water column has an initial volume of 0.05 m$^3$ and the radius of the semi-cylinder is 0.25 m. For the BCGM case, the computational domain has been divided into several subdomains to achieve a high-quality mesh. The cell sizes are similar in each case, resulting in a similar total number of cells up to approximately 0.04 million. 

\begin{figure}[H]
	\centering
	\includegraphics[width=0.9\textwidth]{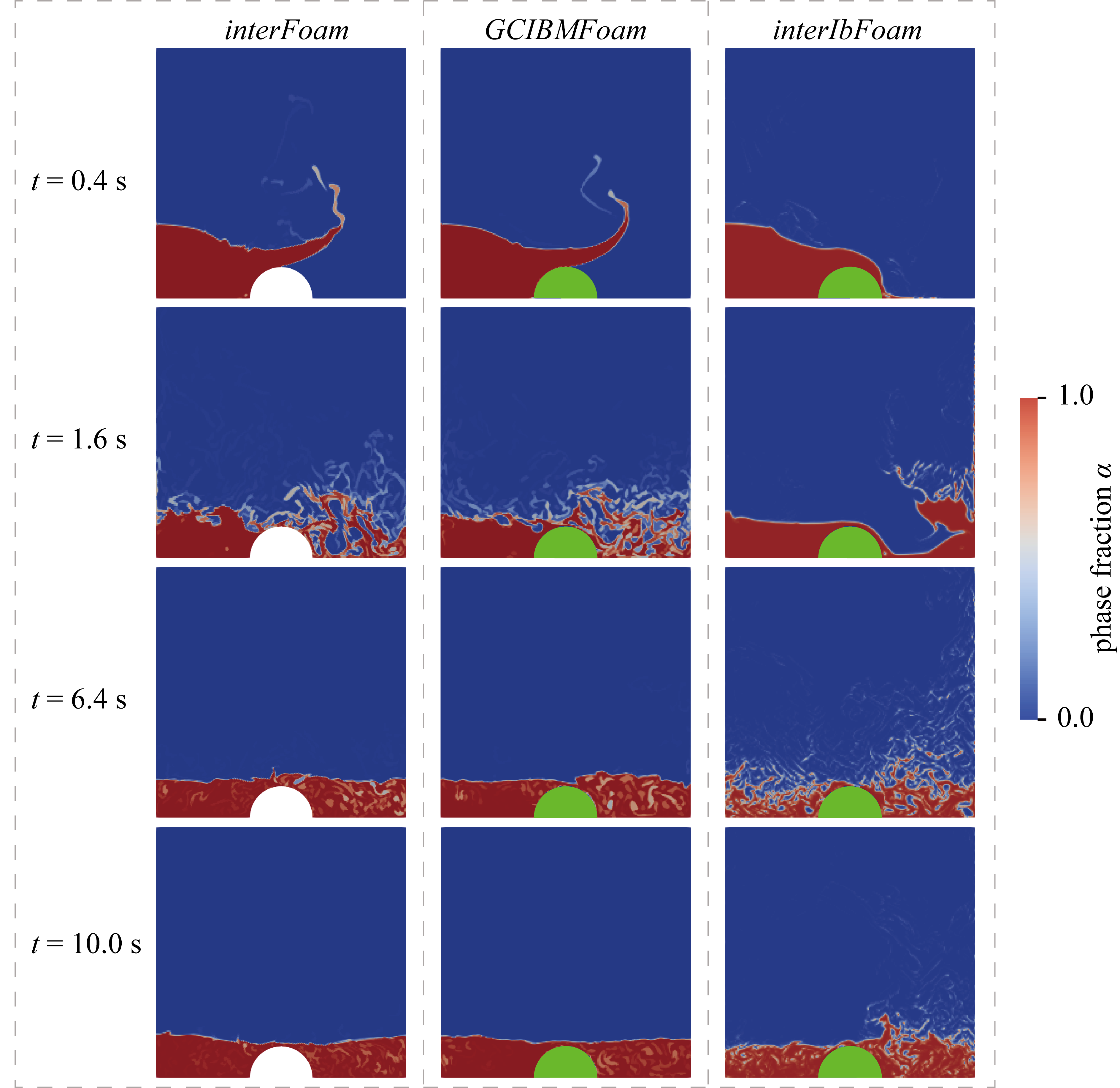}
	\caption{Snapshots of the flow fields simulated by BCGM solver - {\it interFoam}, the newly developed GCIBM solver - {\it GCIBMFoam} and the IBM solver in Foam-extend v4.0 {\it interIbFoam} during the dam break flow impacting on a semi-cylinder structure}
	\label{fig:snap_damBreak}
\end{figure}

The comparison of the flow field evolutions at selected time instants is presented in Fig. \ref{fig:snap_damBreak}. The flow field in the BCGM case generally agrees well with that in the GCIBM case at each time instant. At time {\it t} = 0.4 s, a jet is developed due to the flow impact and the smooth geometry of the structure in both cases. The flow separation also occurs at the top area of the semi-cylinder. However, the above flow phenomena are not observed in the case simulated by the solver involving the disturbance issue, as the water covers the entire semi-cylinder. Based on the results obtained by the BCGM and GCIBM solvers, the splashes are shown on the right side of the domain at {\it t} = 1.6 s due to the violent FSI. With the dissipated energy, the flow fields then tend to become calm as shown in the snapshots at {\it t} = 6.4 and 10.0 s. In comparison, due to the disturbance issue in the {\it interIbFoam} case, the diffused phase fraction is generated around the structure. It affects the subsequent evolution of the flow field, illustrating the non-physical flow phenomenon.

\begin{figure}[H]
	\centering
	\includegraphics[width=0.65\textwidth]{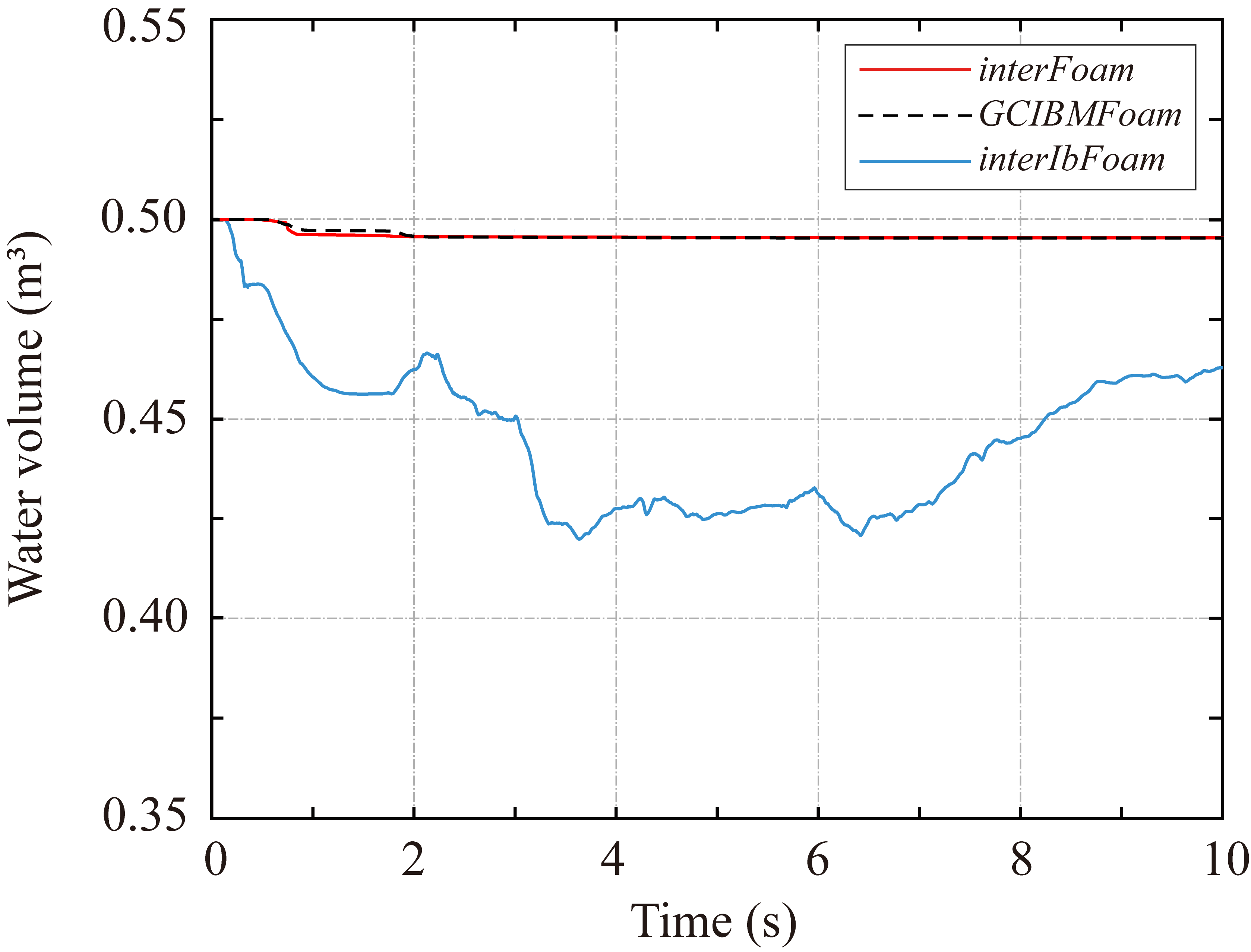}
	\caption{Time history of the water volumes in the simulations by the newly developed solver {\it GCIBMFoam}, original BCGM solver interFoam and the IBM solver {\it interIbFoam} (Foam-extend v4.0)}
	\label{fig:result_damBreak}
\end{figure}

The water volume is measured at each time step. The comparison of the water body volume during the simulation among the different solvers is shown in Fig. \ref{fig:result_damBreak}. Since {\it interIbFoam} has the disturbance issue as shown in Fig. \ref{fig:snap_damBreak}, the total water volume always changes non-physically. In contrast, the flow field obtained by the newly developed GCIBM solver matches very well with that by the BCGM solver, demonstrating its ability to maintain mass conservation and accurately simulate the FSI involving free surface flow. 

For the disturbance issue, the proposed GCIBM, on one hand, avoids directly applying fitted large gradients, which is the important reason for the issue \cite{Jasak_2018}. The fitting method (MLS) is only applied in the initial reconstruction of variables and they are subsequently corrected with $r_{A(u)}$ , which can account for the large differences of variables across the free surface, within the flux correction module and in the process of solving PPE. On the other hand, the proposed GCIBM mimics the BC imposition in BCGM based on the approximated sharp boundary and then considers the non-orthogonal effect and re-evaluates the variable locally to consider the FSI between the actual structure boundary and the surrounding flow field, without introducing further fitting errors. This helps to ensure the accurate capture of FSI as demonstrated by the flow field evolution. Therefore, the proposed GCIBM methodologically circumvents the disturbance issue. The comparison in this subsection also verifies the effective strategy proposed in this work in preserving the fluid mass and well reproducing the interaction between the structure and free surface flow. 

\subsection{Order of accuracy analysis} \label{sec: accuracy}
\begin{figure}[H]
	\centering
	\includegraphics[width=0.6\textwidth]{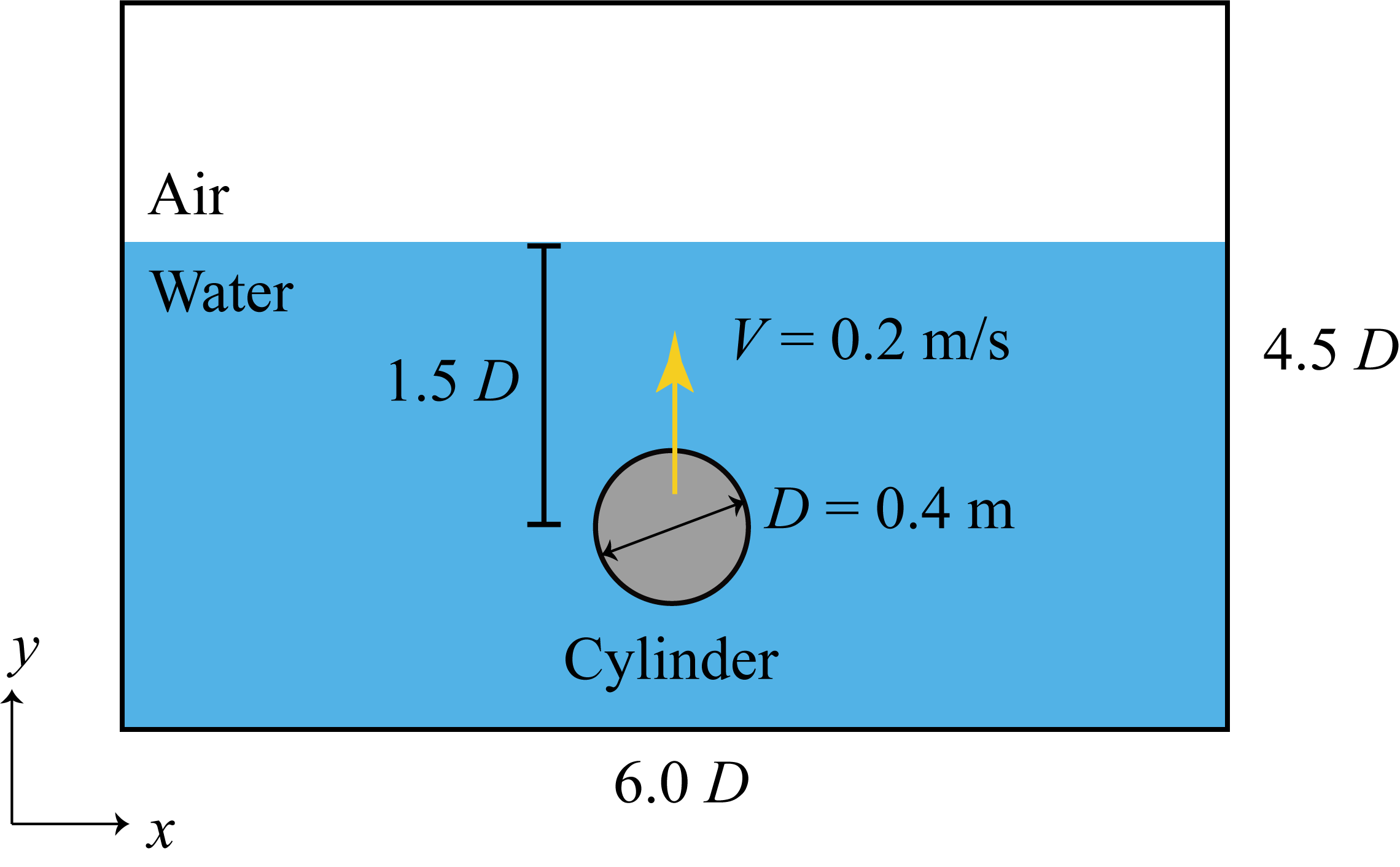}
	\caption{Configuration of the 2D numerical simulation of the water exit of a cylinder}
	\label{fig:setup_waterExit}
\end{figure}

In this work, the temporal term is discretised with first-order accuracy and the discretisation of the remaining terms has second-order accuracy. The overall accuracy of the newly implemented solver is thus expected to be less accurate than the second-order approach. In this section, a 2D water exit of a cylinder is chosen to investigate the accuracy. The numerical setup is illustrated in Fig. \ref{fig:setup_waterExit}, with $D$ denoting the diameter of the cylinder. The cylinder performs a vertical transition with a constant velocity of 0.2 m/s. Unlike steady flow cases used in the order of accuracy analysis, this case involves transient flow and thus the order of accuracy is evaluated at each time step. Three different grid resolutions are selected, namely 200 $\times$ 150, 400 $\times$ 300 and 800 $\times$ 600 cells in the horizontal and vertical directions, respectively. Note that allowing a relatively large Courant number is one of the advantages of PIMPLE, the range of the maximum Courant numbers in this work is 0.3 - 0.9. Additionally, a relaxation factor with a range of 0.5 - 0.8 for dynamic pressure is utilised after each iteration of solving the PPE.

\begin{figure}[H]
	\centering
	\includegraphics[width=0.65\textwidth]{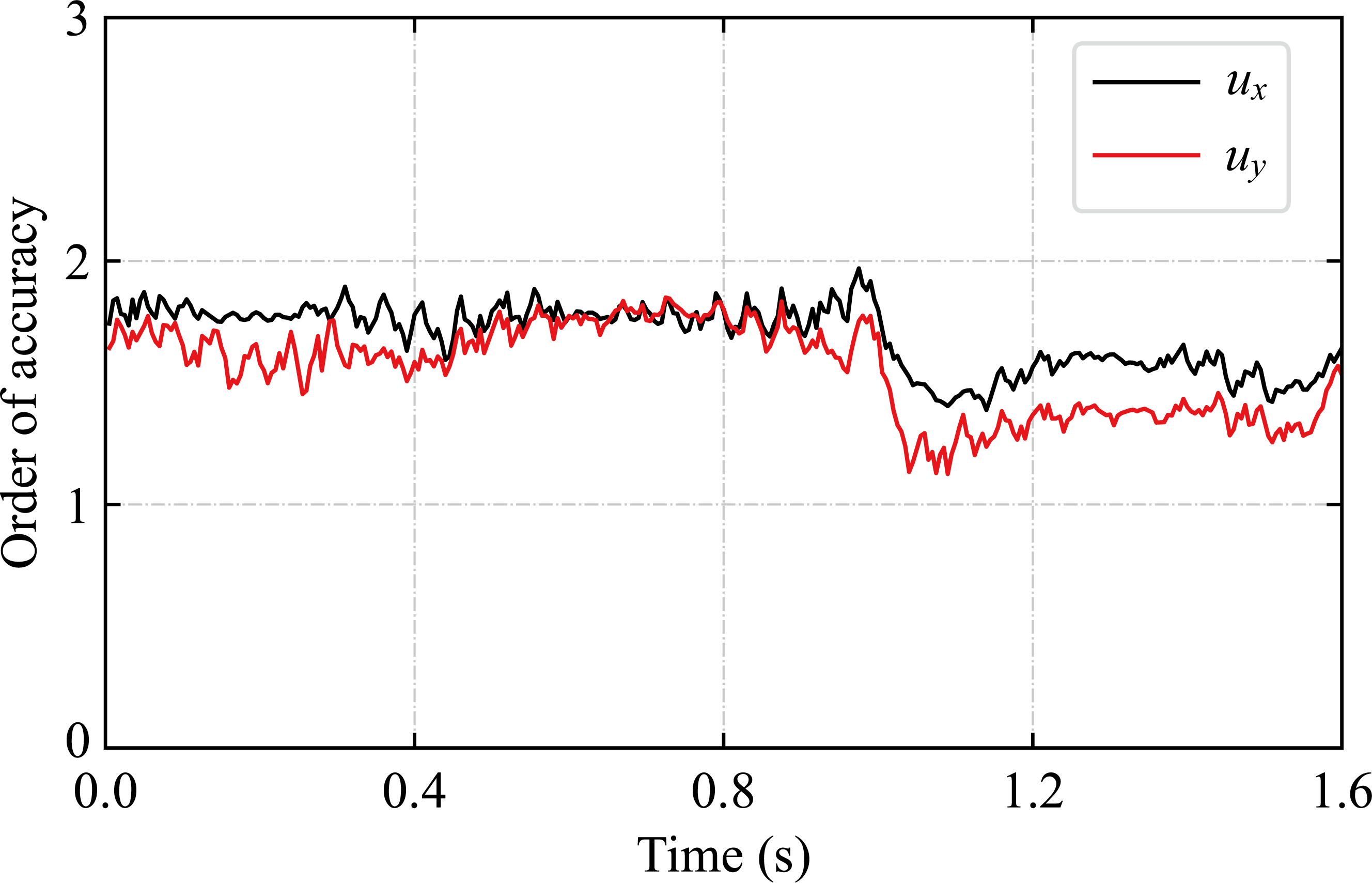}
	\caption{Time series of the evaluated order of accuracy during the water entry process}
	\label{fig:result_waterExit}
\end{figure}

The overall order of accuracy is investigated in this section and it is evaluated according to Kingora and Sadat-Hosseini \cite{KingoraS-H_2022}:
\begin{equation} \label{eqn: orderOfAccuracy}
	n = {{ln(M_i^F-M_i^M) - ln(M_i^M-M_i^C)}\over ln(k_r)},
\end{equation}

\noindent where $M_i=\sum_{\rm j}^{\rm N}|u_{i, {\rm j}}|{\rm\Delta} V_{\rm j}$ is the sum of the i$^{th}$-component of the cell velocity, with N denoting the total number of cells in the computational domain and ${\rm\Delta} V_{\rm j}$ the volume of cell j. The superscripts $F$, $M$ and $C$ indicate the results are obtained by fine, medium and coarse grids, respectively. The grid refinement factor $k_r$ is 2.0 in this case. To eliminate the effect of grid resolution on the evolution of the water surface using VOF, the simulation time window is short such that the cylinder does not move out of the water surface. The results are shown in Fig. \ref{fig:result_waterExit}. As anticipated, the overall order of accuracy remains to be smaller than the second order.

\subsection{Stationary IB case} \label{sec: fixedIB}
\begin{figure}[H]
	\centering
	\includegraphics[width=\textwidth]{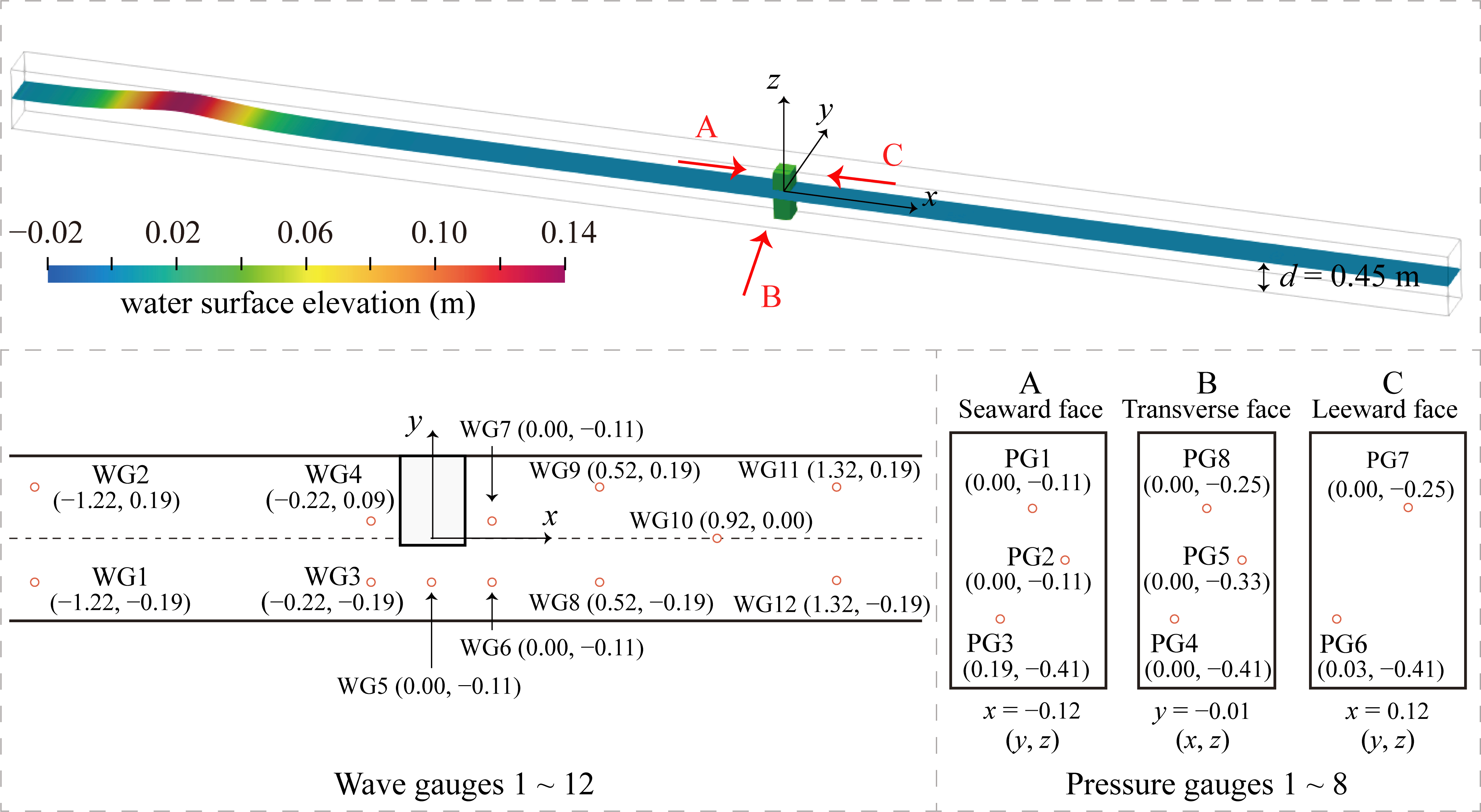}
	\caption{Sketch of the numerical setup and the coordinates of the wave and pressure gauges (adapted from \cite{Higuera_2013}. Dimensions are not scaled)}
	\label{fig:setup_solitaryWave}
\end{figure}

The wave-structure interaction is one of the important topics of free surface flows in engineering problems. In this section, a laboratory test conducted by Higuera et al. \cite{Higuera_2013} is numerically reproduced, where a solitary wave passes over a block. Figure \ref{fig:setup_solitaryWave} shows the corresponding numerical setup. A wave flume with the dimension of 19.80 m $\times$ 0.58 m $\times$ 0.80 m is established, the same as the one without the relaxation zone used in \cite{Bihs_2016}. A Cartesian grid with the areas near the free surface and the block refined is created for the simulation, with the total number of cells up to 6.32 million. The wave generation module originally provided by OpenFOAM is employed for solitary wave generation at the left boundary (Fig. \ref{fig:setup_solitaryWave}). The working principle of this wave generation boundary condition is to specify the velocity and the phase fraction at each time step based on the Boussinesq-type solitary wave theory \cite{DeanDalrymple_1991}. For a given wave amplitude A and water depth $d$, the water surface elevation $\eta$ and the wave celerity $c$ are obtained as
\begin{equation} \label{eqn: eta}
	\eta = A{sech}^2\left[\sqrt{3A\over4d^3}(x-ct)\right],
\end{equation}
\begin{equation} \label{eqn: celerity}
	c = \sqrt{\left|{\bf g}d\right|\over{1-{A\over d}}}.
\end{equation}

\noindent According to the experimental setup in \cite{Higuera_2013}, $A$ = 0.14 m and $d$ = 0.45 m. The water surface elevations and dynamic pressures are measured at the same locations as those measured in the laboratory tests as shown in Fig. \ref{fig:setup_solitaryWave}. The simulation is run in parallel mode (9 cores), taking about 70.0 hours to simulate 15.0 s in real time.

\begin{figure}[H]
	\centering
	\includegraphics[width=0.9\textwidth]{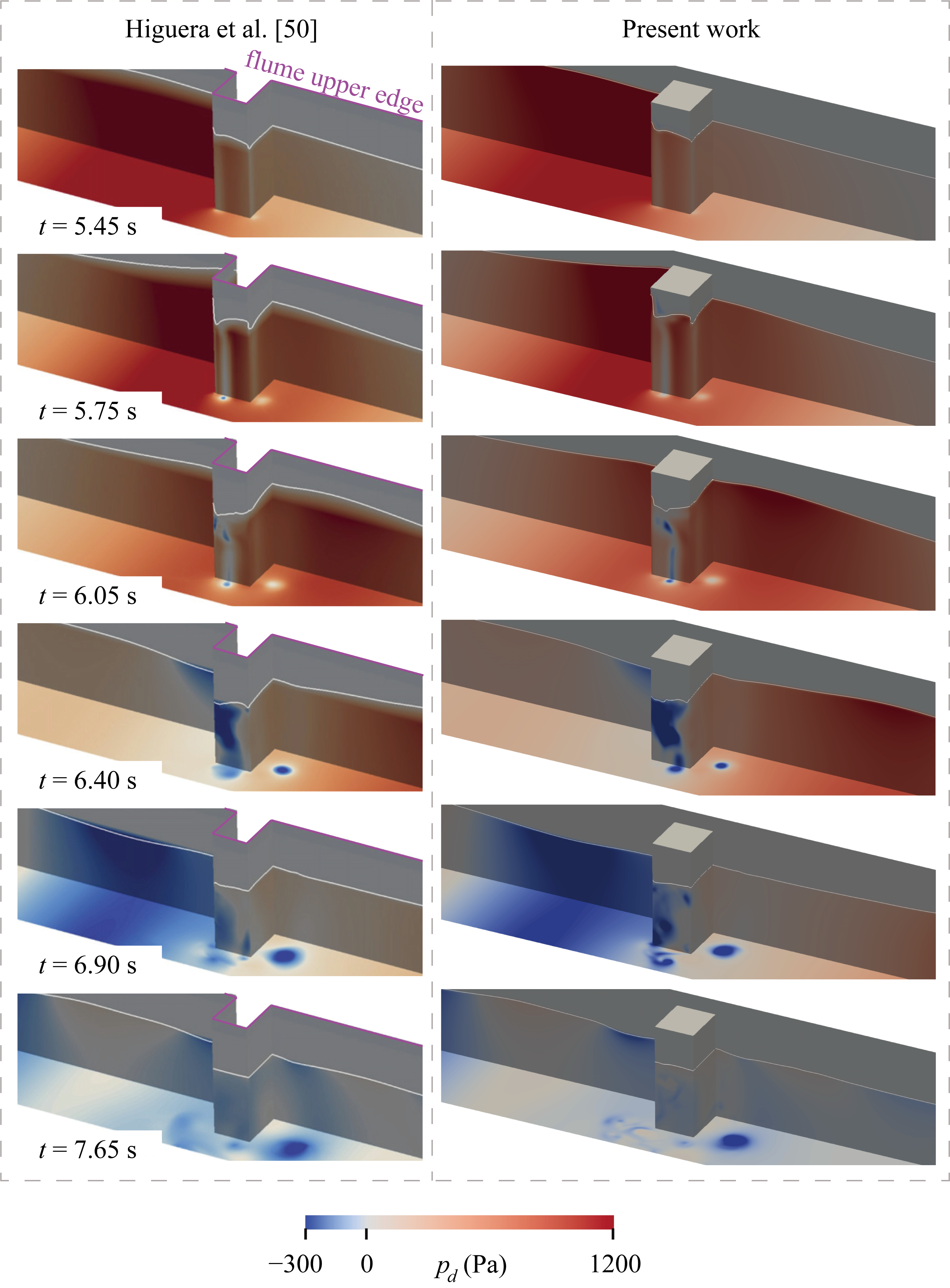}
	\caption{Snapshots of dynamic pressure together with the water surface contour (white solid line) on the boundary at different time instants (adapted from \cite{Higuera_2013})}
	\label{fig:snap_solitaryWave}
\end{figure}

The evolution of dynamic pressure together with the free surface contour on the flume and block surface are presented in Fig. \ref{fig:snap_solitaryWave}. During the solitary wave interacting with the block, the solitary wave generally results in positive dynamic pressure in both the flow fields simulated by the BCGM \cite{Higuera_2013} and the present GCIBM solvers. Due to the presence of the block affecting the free surface, negative dynamic pressure can be observed mainly on the flume bottom and the block’s transverse face. A circular zone with pressure drop on the flume bottom is generated since $t$ = 6.40 s in both simulations, which then continues to develop. It should be noted that there is a good agreement on the increasing area of this pressure drop between the numerical simulations in \cite{Higuera_2013} and the present work, demonstrating the similar evolution of $p_d$ and the free surface. By comparison of the flow field evolution with the BCGM solver, it further shows the aforementioned disturbance has been effectively suppressed.

\begin{figure}[H]
	\centering
	\includegraphics[width=0.97\textwidth]{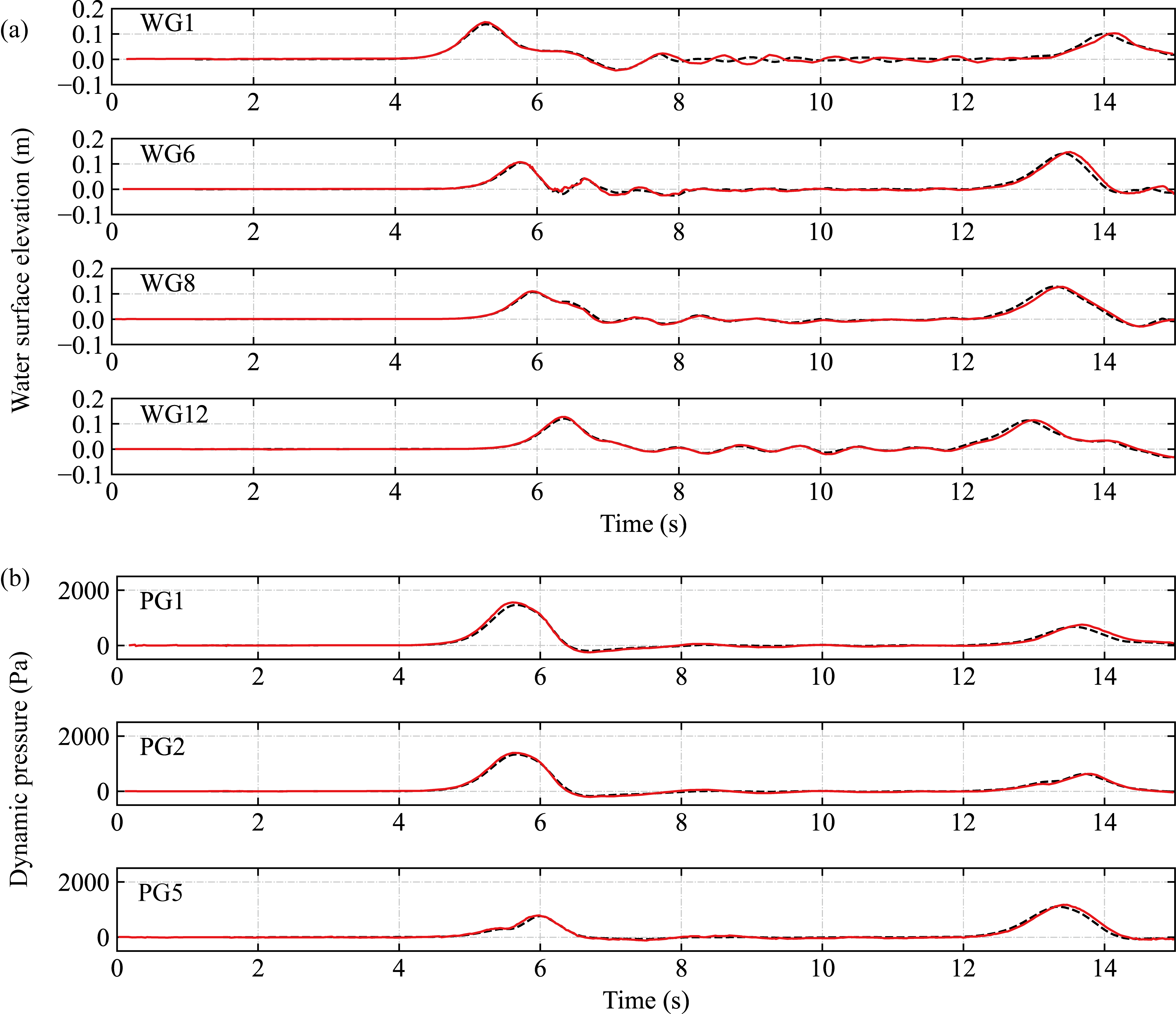}
	\caption{The comparison of water surface elevations and dynamic pressures between the numerical simulation (black dash lines) and laboratory tests (red solid lines) in \cite{Higuera_2013}}
	\label{fig:result_solitaryWave}
\end{figure}

The time series of the water surface elevation and dynamic water pressure is presented in Fig. \ref{fig:result_solitaryWave}, together with the results from Higuera et al. \cite{Higuera_2013}. Note that half of the entire solitary waveform is already initialised in the pre-processing stage, which differs from the laboratory tests. The time series are thus shifted to account for the time difference (about 1.17 s). The numerical results match well with the laboratory test data, especially regarding water surface elevations. The peak dynamic pressures are slightly underpredicted compared with the laboratory results, with a maximum deviation of 6.5\%. Nevertheless, the time instants corresponding to the pressure peak and trending are well reproduced. This validation case illustrates the well-captured FSI by the developed solver.

\subsection{Moving IB cases} \label{sec: movingIB}
\subsubsection{Water entry of a free-falling wedge} \label{sec: waterEntry}
\begin{figure}[H]
	\centering
	\includegraphics[width=0.75\textwidth]{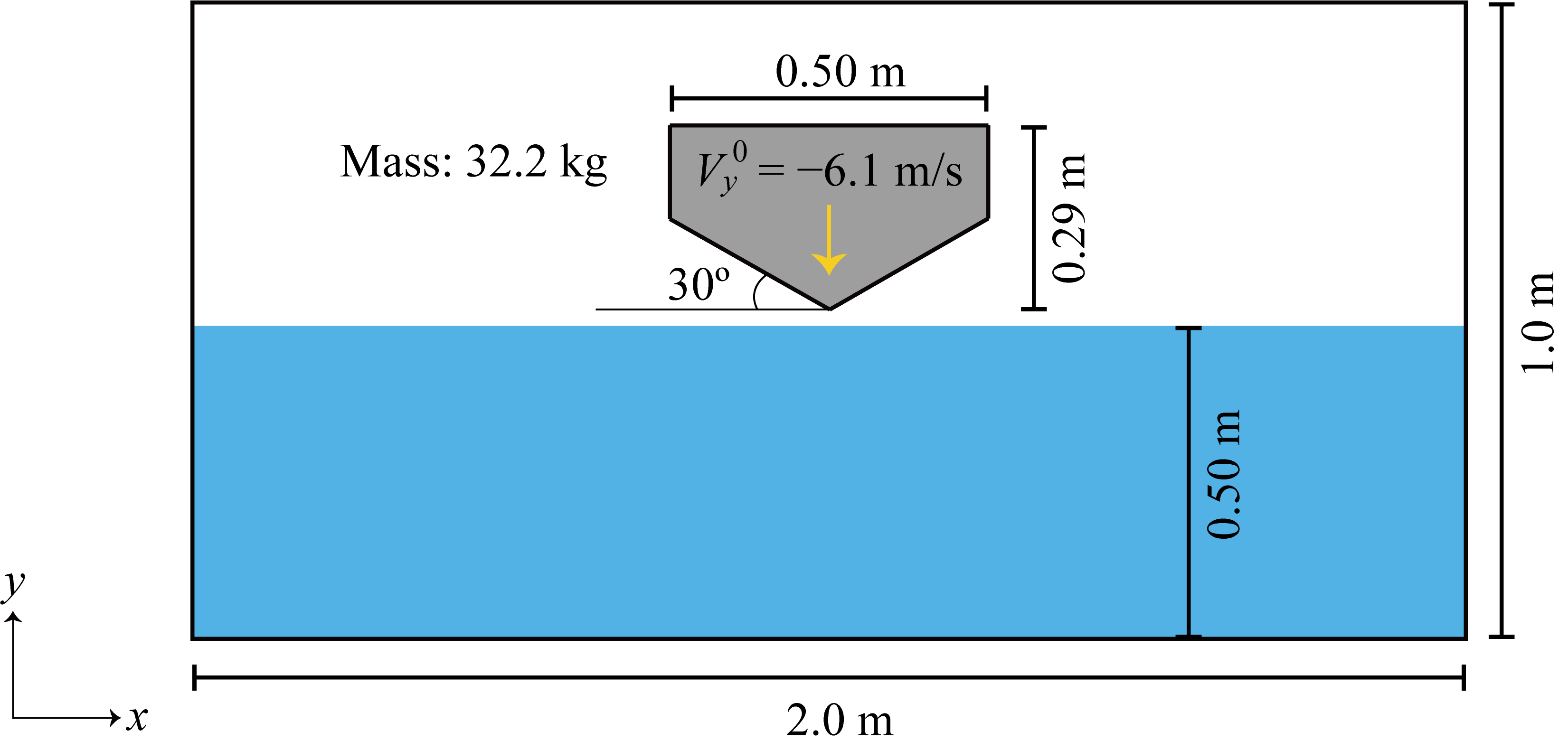}
	\caption{Sketch of the numerical setup based on the drop test from Zhao et al. \cite{Zhao_1997}}
	\label{fig:setup_waterEntry}
\end{figure}

This section reproduces a case of a moving IB with resolved motion, which is different from the water entry simulation using the prescribed motion in Section 3.1. This numerical simulation setup is based on an experimental test conducted by Zhao et al. \cite{Zhao_1997}, in which a wedge was dropped freely into a calm water body. In this drop test, the symmetry wedge has dimensions of 0.50 m $\times$ 0.29 m $\times$ 1.00 m with a deadrise angle of 30$^{\circ}$, as shown in Fig. \ref{fig:setup_waterEntry}. The total mass of the wedge is up to 161.0 kg. Since this is a 2D numerical simulation, the wedge length along the $z$-axis is reduced to 0.2 m, resulting in a mass of 32.2 kg. The same cell size used in other numerical simulations (0.0025 m $\times$ 0.0025 m) is also applied in this simulation. The time instant $t$ = 0.0s corresponds to when the wedge starts interacting with the water surface.

\begin{figure}[H]
	\centering
	\includegraphics[width=0.85\textwidth]{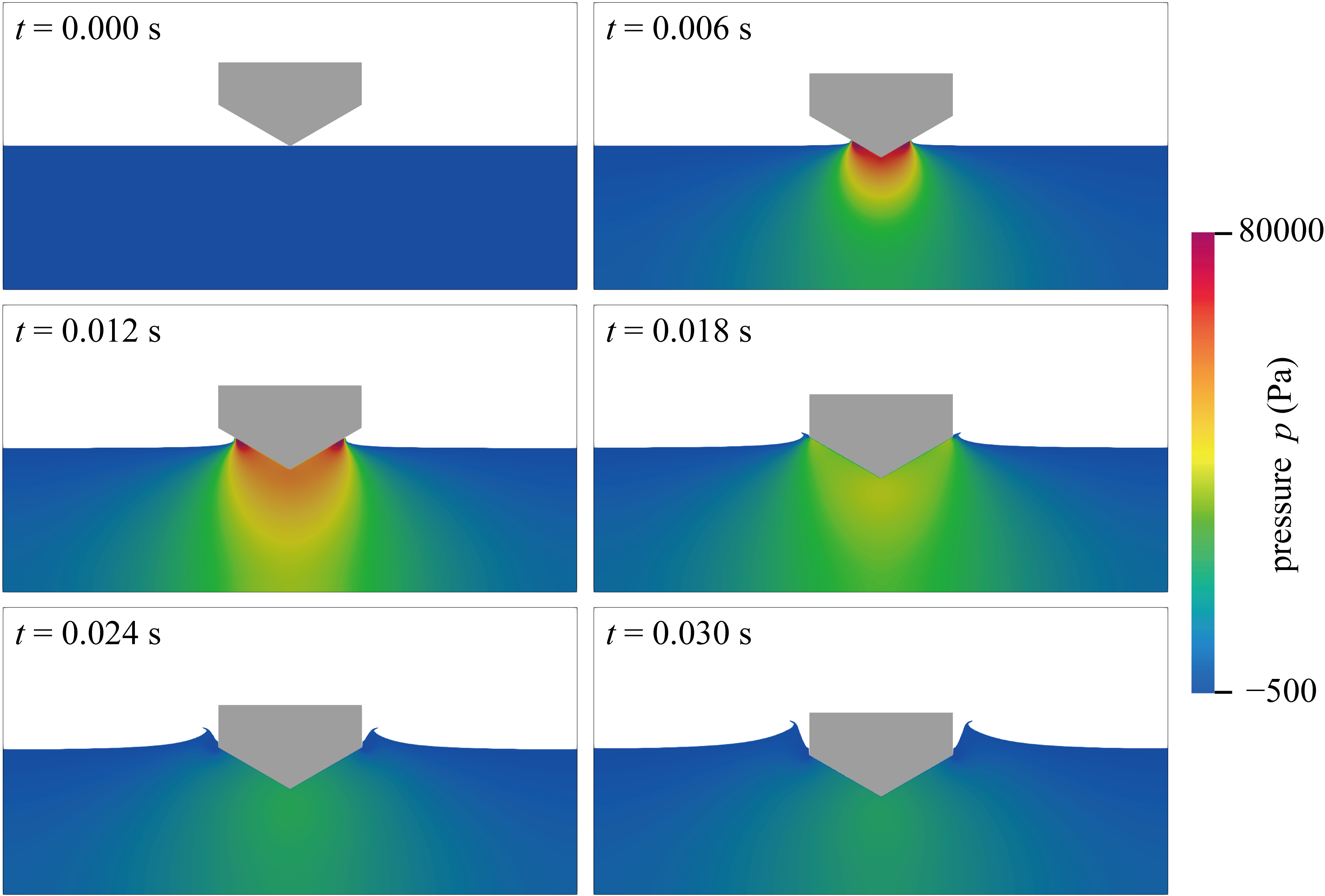}
	\caption{Evolution of water pressure at selected time instants}
	\label{fig:snap_waterEntry}
\end{figure}

Figure \ref{fig:snap_waterEntry} shows the snapshot series of the water pressure. With the continued impact of the structure, the water surface attaches to the edges of the wedge part. The thin jet flow is generated at each side of the wedge at $t$ = 0.030 s. In the high impact zone, it can be seen the pressure is smoothly varied without disturbance occurring around the structure. Further, during the water entry process, large differences in pressure and phase fraction across the free surface can be clearly observed, without diffusions. This again indicates the effectiveness of the proposed strategy in suppressing the disturbance. 

\begin{figure}[H]
	\centering
	\includegraphics[width=0.8\textwidth]{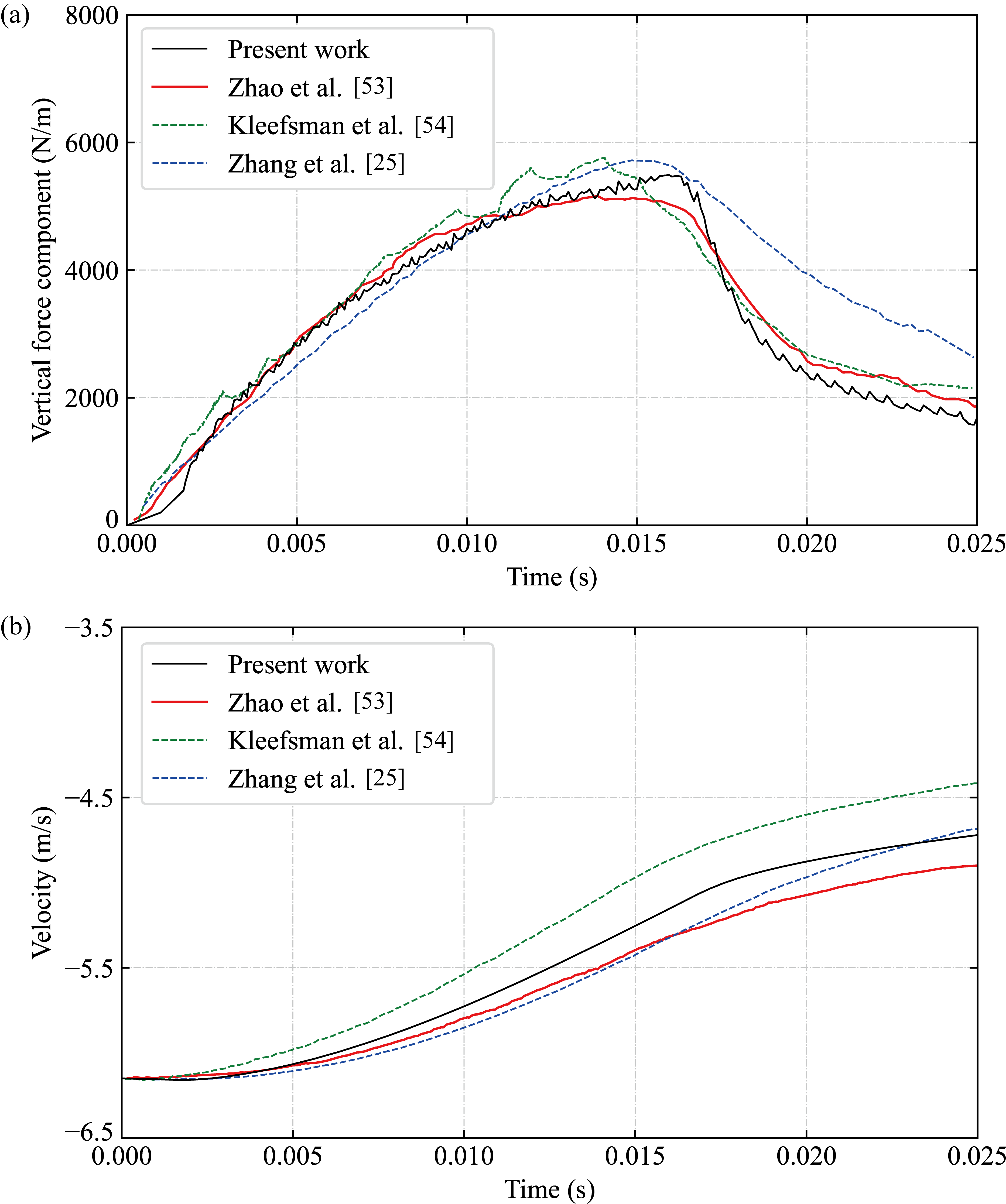}
	\caption{Comparison of (a) forces acting on the falling wedge and (b) wedge velocity among the laboratory data from \cite{Zhao_1997}, the present and other numerical simulation results in \cite{Zhang_2010, Kleefsman_2005}}
	\label{fig:result_waterEntry}
\end{figure}

The numerical results from \cite{Zhang_2010} by GCIBM and \cite{Kleefsman_2005} via cut-cell IBM are used for the comparison. Figure \ref{fig:result_waterEntry}(a) shows the time history of the pressure force exerted on the wedge. The numerical results generally agree well with those in both the laboratory tests and other numerical simulations. The peak pressure is overpredicted by approximately 6.4\% when compared with the laboratory data, although some non-physical pressure oscillations with small amplitudes can be observed in the curve. This is because the source term correction in PPE does not account for the changed cell volume based on the projected non-orthogonal IB. Nevertheless, compared with other numerical studies, there is a better fit between the present numerical and laboratory peak pressure. In addition to the well-matched maximum pressure force, the current numerical results show almost the same trend of the pressure as \cite{Zhao_1997} and \cite{Kleefsman_2005}, leading to similar pressure (approximately 1.85 $\times$ 103 N/m) at $t$ = 0.025 s as measured in the laboratory test. The comparison of the wedge velocity among the present study, numerical results from \cite{Zhang_2010, Kleefsman_2005} and laboratory measurement by \cite{Zhao_1997} are shown in Fig. \ref{fig:result_waterEntry}(b). The present numerical reproduced motion along with that from \cite{Zhang_2010} agrees better with the laboratory measurement. At $t$ = 0.025 s, the present numerical simulation has the best agreement with the laboratory measurement among the numerical results, with a relative error of 3.6\%. This validation case demonstrates the accurate simulation of FSIs involving the resolved motion of the structure.

\newpage  
\subsubsection{Simulation of tsunamis generated by iceberg calving}  \label{sec: IBT}
\begin{figure}[H]
	\centering
	\includegraphics[width=\textwidth]{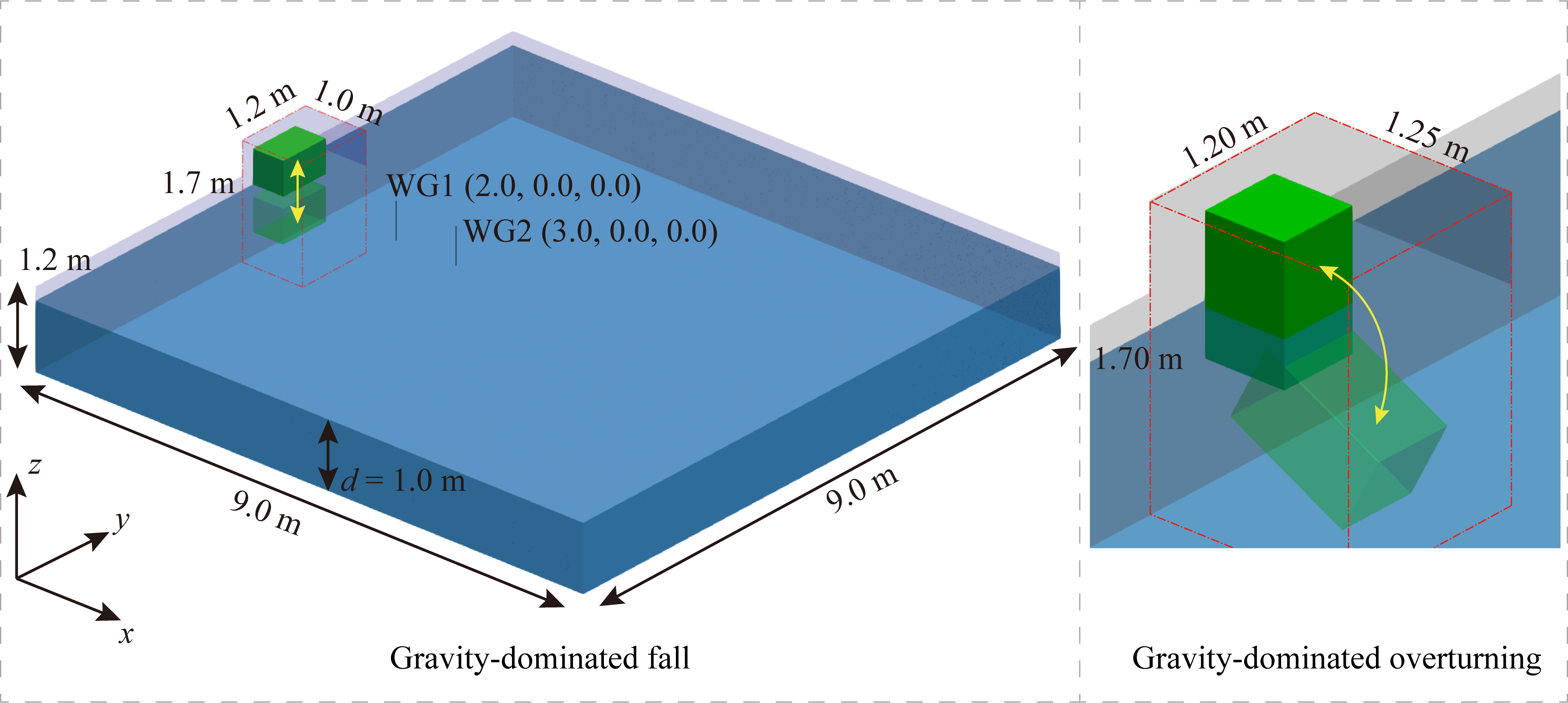}
	\caption{Numerical wave basins for the gravity-dominated fall and overturning cases}
	\label{fig:setup_IBT}
\end{figure}

To examine the flow field evolution involving moving IBs, the impulse waves generated by iceberg calving (hereafter referred to as iceberg-tsunami, short as IBT) are simulated and the results are compared with those of \cite{Heller_2019}. Given that the gravity-dominated cases of iceberg calving are threatening, the fall and overturning cases are simulated in this work. The numerical setup is based on the laboratory tests presented in \cite{Heller_2019}. A smaller wave basin with the dimension of 9.0 m $\times$ 9.0 m $\times$ 1.2 - 1.7 m is created in this work in order to compare the generated leading waves. Two wave gauges (WG1 at $x$ = 2.0 m and WG2 at $x$ = 3.0 m) are used to measure the water surface elevation such that the leading wave will not be affected by the reflection from side walls. The same coordinate as that in \cite{Chen_2020} is established as shown in Fig. \ref{fig:setup_IBT}. Grid refinement is made within the wave generation zone (outlined with red dash lines in Fig. \ref{fig:setup_IBT}). In the gravity-dominated overturning case, there is a gap between the block and the boundary to facilitate the evaluation using MLS. A gap with a width of 0.05 m is chosen and grid refinement along the $x$-axis direction is performed in this gap zone ($-$0.05 $\le x \le$ 0.00). Both cases have a total number of about 6.9 million cells in the computational domain. The iceberg performs translation and rotation in these two cases, respectively, and the motion is resolved based on the flow field at each time step. The parallel running with 12 cores is applied, taking approximately 30.0 hours to simulate 6.0 s.

\begin{figure}[H]
	\centering
	\includegraphics[width=\textwidth]{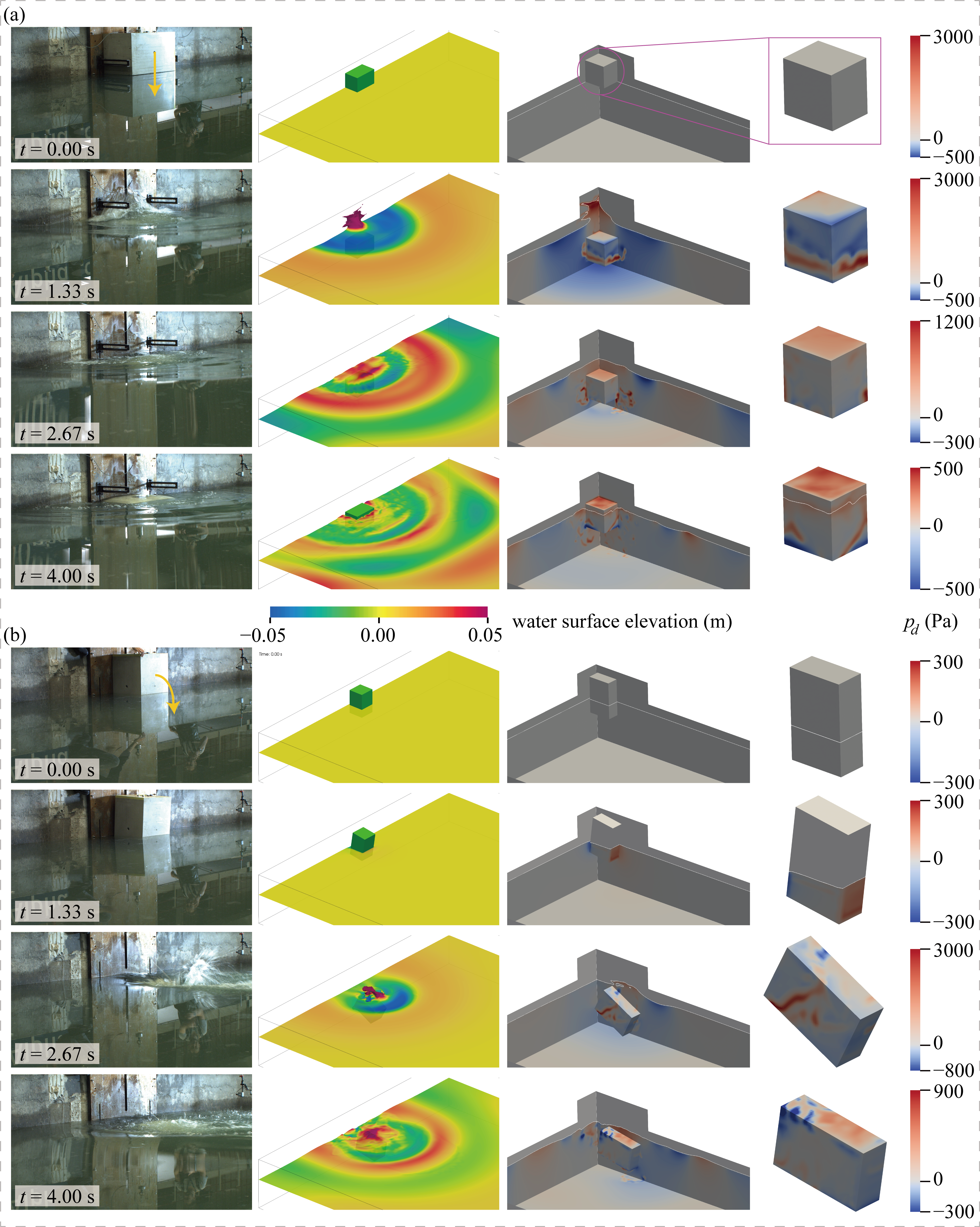}
	\caption{Snapshots of numerical simulations and their corresponding laboratory photos in (a) gravity-dominated fall and (b) gravity-dominated overturning cases (adapted from \cite{Chen_2020})}
	\label{fig:snap_IBT}
\end{figure}

The snapshots series of the flow fields at selected time instants are shown in Fig. \ref{fig:snap_IBT}, including the water surface elevation and the dynamic pressure in the half domain. The numerical simulations successfully reproduce the IBTs in both cases. The locations of icebergs generally agree well with those in the laboratory tests at different time instants. After the icebergs impact the water body, waves start to propagate in a semi-circle pattern until the waves are reflected by the side walls. Meanwhile, the dynamic pressure evolution also agrees with the slamming process. The upper part of the splash and block bottom undergo relatively large pressure at $t$ = 1.33 s in the fall and $t$ = 2.66 s in the overturning cases, respectively. With the block velocity decreasing, its impact on the water body is then obviously alleviated. 

\begin{figure}[H]
	\centering
	\includegraphics[width=0.97\textwidth]{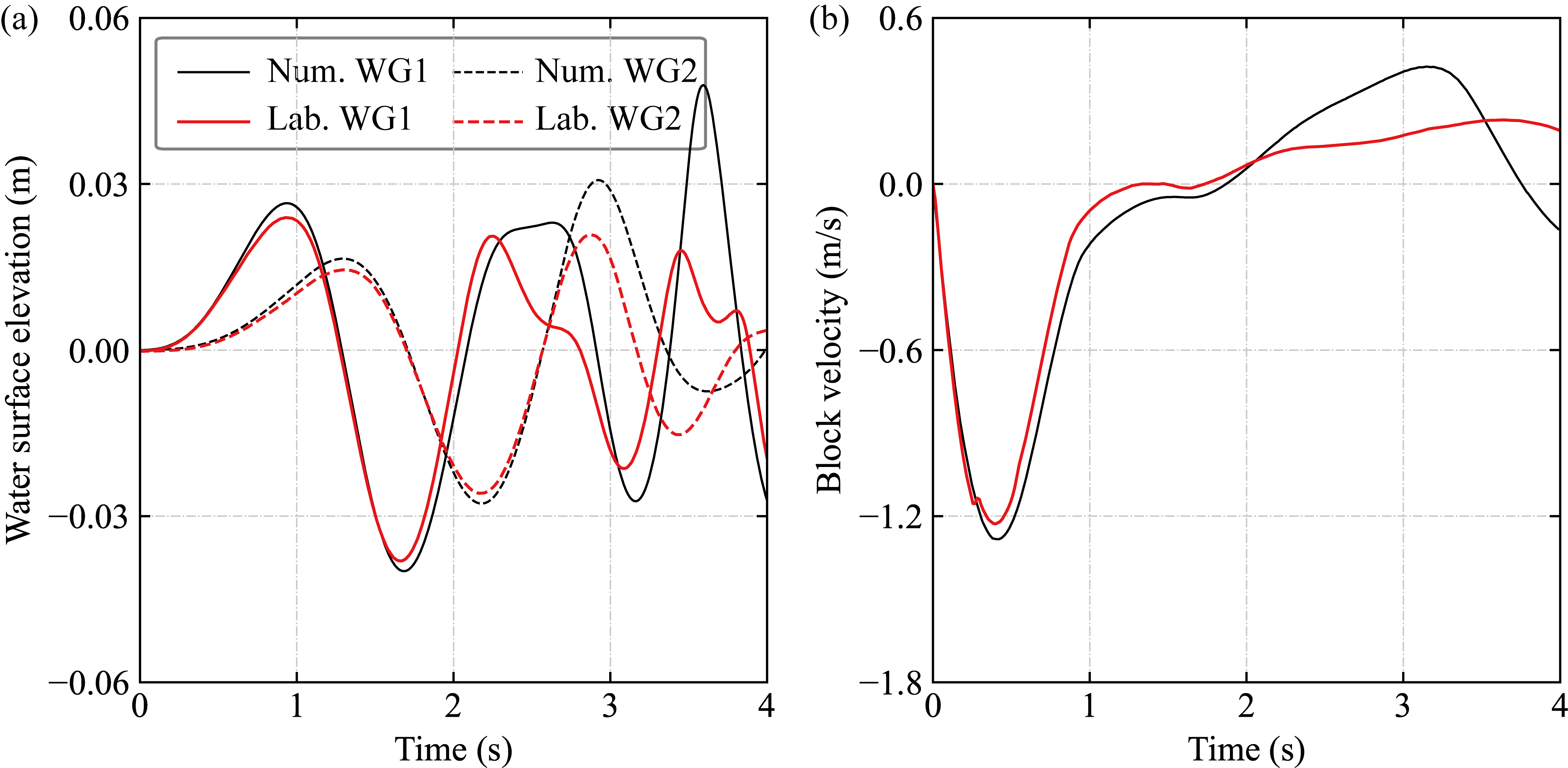}
	\caption{Comparison of (a) the measured waves and (b) block velocity between the present study (black lines) and the laboratory results (red lines) in the gravity-dominated fall case}
	\label{fig:result_IBT_t21}
\end{figure}

\begin{figure}[H]
	\centering
	\includegraphics[width=0.97\textwidth]{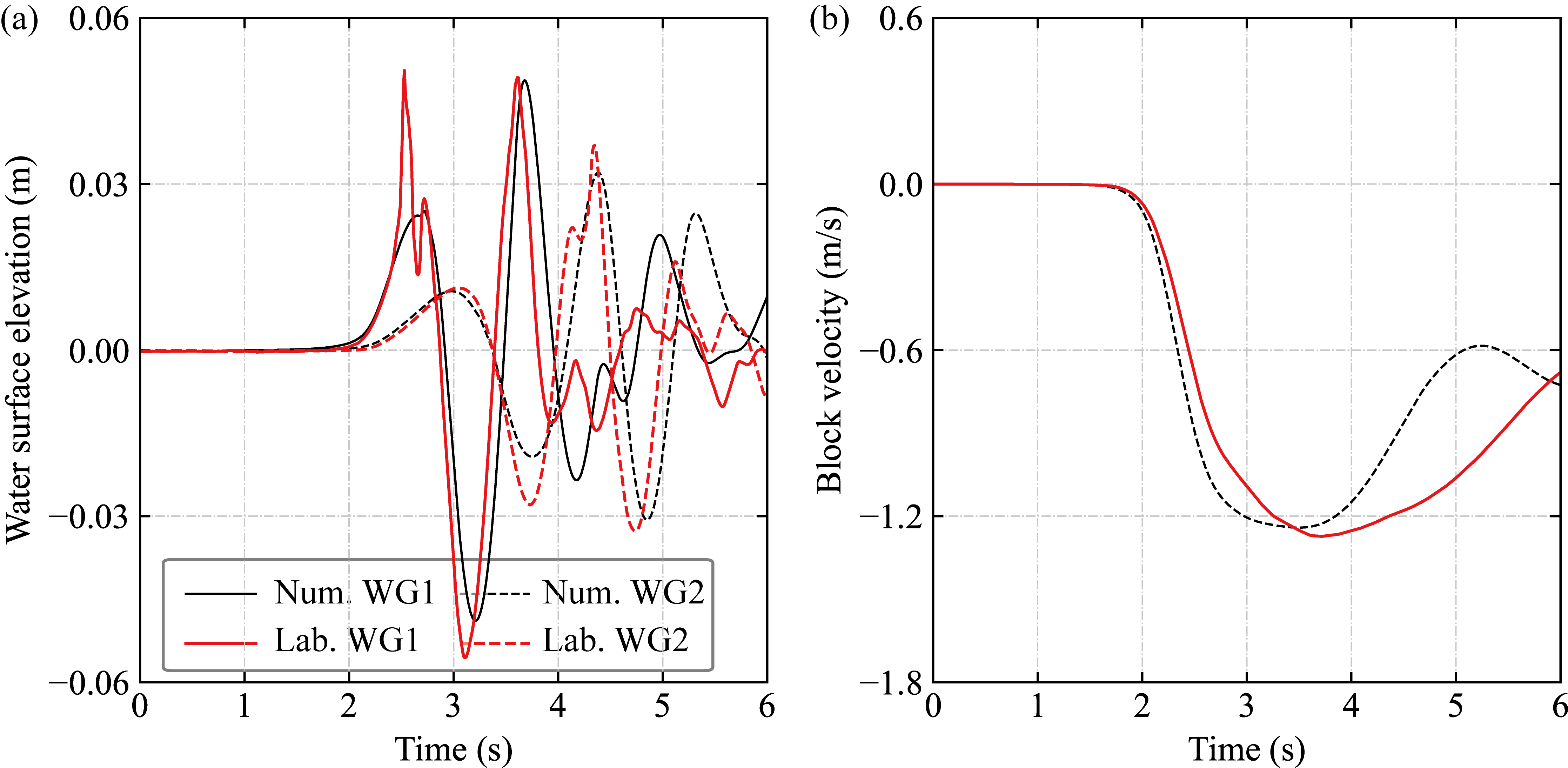}
	\caption{Comparison of (a) the measured waves and (b) block velocity between the present study (black lines) and the laboratory results (red lines) in the gravity-dominated overturning case}
	\label{fig:result_IBT_t24}
\end{figure}

Figure \ref{fig:result_IBT_t21} shows the comparisons of water surface elevations and iceberg velocity between the simulated fall case and laboratory test. It shows that the newly developed solver is accurate in reproducing the iceberg motion, with the maximum velocity overpredicted by 4.3\% relative to the laboratory measurement. Further, the generated wave has a good agreement with that in the laboratory tests with a deviation of 10.9\% on the amplitude of the leading wave. The time history of the water surface elevation of IBT in the overturning case is shown in Fig. \ref{fig:result_IBT_t24}, together with the laboratory measurement. The numerical leading wave crest and trough are 0.0251 m and $-$0.0489 m, respectively. Since the wave crest was affected by the splash in the laboratory test, the numerical wave crest is nearly half of that in the laboratory measurement. Nevertheless, the first wave trough is well simulated as well as the subsequent wave crest at WG1. The deviations of them are $-$11.9\% and 2.0\%, respectively. The numerically reproduced motion also reasonably agrees with that in the laboratory test, with a slight underestimation of 2.5\% in the maximum vertical velocity component. 

\newpage  
With the stationary and moving IB cases presented in Sections \ref{sec: fixedIB} and \ref{sec: movingIB}, the validation results can be summarised as follows. Firstly, the numerical results show good accuracy of the newly developed solver in simulating free surface flow interacting with structures, demonstrating the successful coupling of GCIBM and PIMPLE. Meanwhile, a relatively large Courant number is allowed, which can save computational time. Secondly, the flow fields are in agreement with those in the corresponding laboratory measurements and numerical simulations, indicating that the disturbance issue has been well avoided. Therefore, fluid mass conservation is well preserved. For moving IB cases, despite the smoothing process introduced in this work to deal with the changed cell identities, some pressure oscillations can still be observed in the moving IB cases. This is due that the calculation of flux based on the projected non-orthogonal IB may not fully account for the actual structure geometry details. A more robust and accurate way to calculate the scaled flux and the changed cell volume in the PPE’s source term may be helpful for further suppressing the pressure oscillation.

\section{Conclusion} \label{sec: conclusions}
This article presents a novel volume of fluid Ghost-cell Immersed Boundary Method (GCIBM) for two-phase free surface flow (i.e., the air and water) interacting with structures. To avoid the disturbance around the intersection area of the immersed boundary (IB) and free surface, the Fluid-Structure Interaction (FSI) is initially considered on the “stair-step” (orthogonal) IB, by mimicking the boundary condition imposition in the body-conformal grid method. After this, in order to accurately simulate FSI with the consideration of the actual structure boundary, a flux scaling method is proposed to account for the non-orthogonal effect and the IB velocity is re-evaluated accordingly based on the non-orthogonal IB. When dealing with the moving IB cases, a variable smoothing process is performed to address unphysical values due to identity-changed cells. Further, the flux correction is introduced to avoid the fake velocity near the IB for the two-phase flow with a large density ratio. A two-phase flow solver has been developed based on the open-source computational fluid dynamics code OpenFOAM. Modules including the force calculation and moving IB handling, have also been developed to support the simulation of FSI involving free surface flow.

The developed solver is first verified regarding the circumvented disturbance issue. The order of accuracy of this solver is then evaluated and an expected accuracy of less than second order is obtained. Through the validations with canonical cases involving both the stationary and moving IBs, reasonably good agreements are obtained in comparison with corresponding laboratory data and numerical results, demonstrating the effectiveness of the coupling of GCIBM and PIMPLE in dealing with free surface flow. Moreover, the disturbance effect has been effectively suppressed so that the FSI can be accurately simulated with fluid mass well preserved. 

Regarding the handling of moving IB cases, small amplitude pressure oscillations can be observed. This may be due that the approximated source term in the Poisson pressure equation does not fully account for changes in the flux and cell volume caused by the moving IB, which requires further investigations. Future studies will focus on increasing the order of accuracy by potentially involving the Crank-Nicolson scheme for time advancing. An automatic mesh refinement module along with its variable reconstruction method will be introduced for refined simulations of FSI problems in coastal and ocean engineering.

\section*{Acknowledgement}
This work was supported by the National Natural Science Foundation of China (No. 52101366). FC is grateful for the financial support via the Postdoc Matching Funds of The Hong Kong Polytechnic University (project no. 1-W350). The support from the Hong Kong Research Grants Council (RGC, project no. C5002-22Y) is acknowledged.

\section*{Appendix A: Variable Reconstruction using the Moving Least Squares method}\label{appA}
\setcounter{equation}{0}
\setcounter{figure}{0}
\renewcommand{\theequation}{A.\arabic{equation}}
\renewcommand{\thefigure}{A.\arabic{figure}}

The reconstruction using MLS relies on an interpolation stencil, where the value of each stencil point is used in the fitting. The determination of the stencil mainly depends on the distance of a cell relative to the IB cell. For an IB cell variable $\phi_{\scaleto{IB}{4.5pt}}$, it is evaluated by
\begin{equation} \label{eqn: MLS_A}
	\phi_{\scaleto{IB}{4.5pt}} = c_0 + \sum\limits_{\rm k=1}^{\rm N_c}c_{\rm k}p_{\rm k} = {\bf P}^{\rm T}{\bf c}.
\end{equation}

\noindent In Eq. (\ref{eqn: MLS_A}), the basis functions $p_{\rm k}$ are a series of polynomial functions related to the position vector of a stencil cell centre to the IB cell’s centre, i.e., ${\bf X}_c^{stencil}-
{\bf X}_{\scaleto{IB}{4.5pt}}$. With the position vector’s components along each axis denoted as ${\rm\Delta} x$, ${\rm\Delta} y$ and ${\rm\Delta} z$, $p_0=1$, $p_1={\rm\Delta} x$, $p_2={\rm\Delta} y$, $p_3={\rm\Delta} z$, $p_4={\rm\Delta} x{\rm\Delta} y$, $p_5={\rm\Delta} y{\rm\Delta} z$, $p_6={\rm\Delta} x{\rm\Delta} z$, $p_7=({\rm\Delta} x)^2$, $p_8=({\rm\Delta} y)^2$ and $p_9=({\rm\Delta} z)^2$ 
in 3D cases. In 2D cases, the above $p_{\rm k}$ related to z-axis components are neglected. The column vectors {\bf P} and {\bf c} contain $p_{\rm k}$ and $c_{\rm k}$, respectively. The MLS method in this work results in second order of accuracy. 

To obtain $c_{\rm k}$, the distance-based weighting coefficient $w_{\rm i,j}^{MLS}$ is introduced to consider the contributions of each stencil cell on the fitting, which is given as
\begin{equation} \label{eqn: weightingFunc}
	w_{\rm i, j}^{MLS} =
	\begin{cases}
		1 - 6d^2 + 6d^3\quad &{\rm if}\ d/d_{max} \le 0.5\\
		2 - 6d + 6d^2 - 2d^3\quad &{\rm if}\ 0.5 < d/d_{max} \le 1.0\\
		0 &{\rm if}\ d/d_{max} > 1.0
	\end{cases}
\end{equation}

\noindent By introducing the fitting error $J$ and minimising its value using the partial derivative with respect to each interpolation coefficient $\partial J\over\partial c_{\rm k}$, the interpolation coefficients $c_{\rm k}$ should satisfy the following equations
\begin{equation} \label{eqn: MLS_error}
	{\partial J\over\partial c_{\rm k}} = 
	\sum\limits_{\rm j=1}^{\rm N_p}w_{\rm i, j}^{MLS}
	\left(c_0 + \sum\limits_{\rm n=1}^{\rm N_c}c_{\rm n}p_{\rm n} - \phi_{\rm j}
	\right)p_{\rm k} = 0.
\end{equation}

\noindent After rearranging, Eq. (\ref{eqn: MLS_error}) becomes
\begin{equation} \label{eqn: MLS_error_matrix}
	\sum\limits_{\rm j=1}^{\rm N_p}w_{\rm i, j}^{MLS}
	\left(c_0 + \sum\limits_{\rm n=1}^{\rm N_c}c_{\rm n}p_{\rm n} 
	\right)p_{\rm k} = \sum\limits_{\rm j=1}^{\rm N_p}w_{\rm i, j}^{MLS}\phi_{\rm j}p_{\rm k}.
\end{equation}

\begin{figure}[H]
	\centering
	\includegraphics[width=0.85\textwidth]{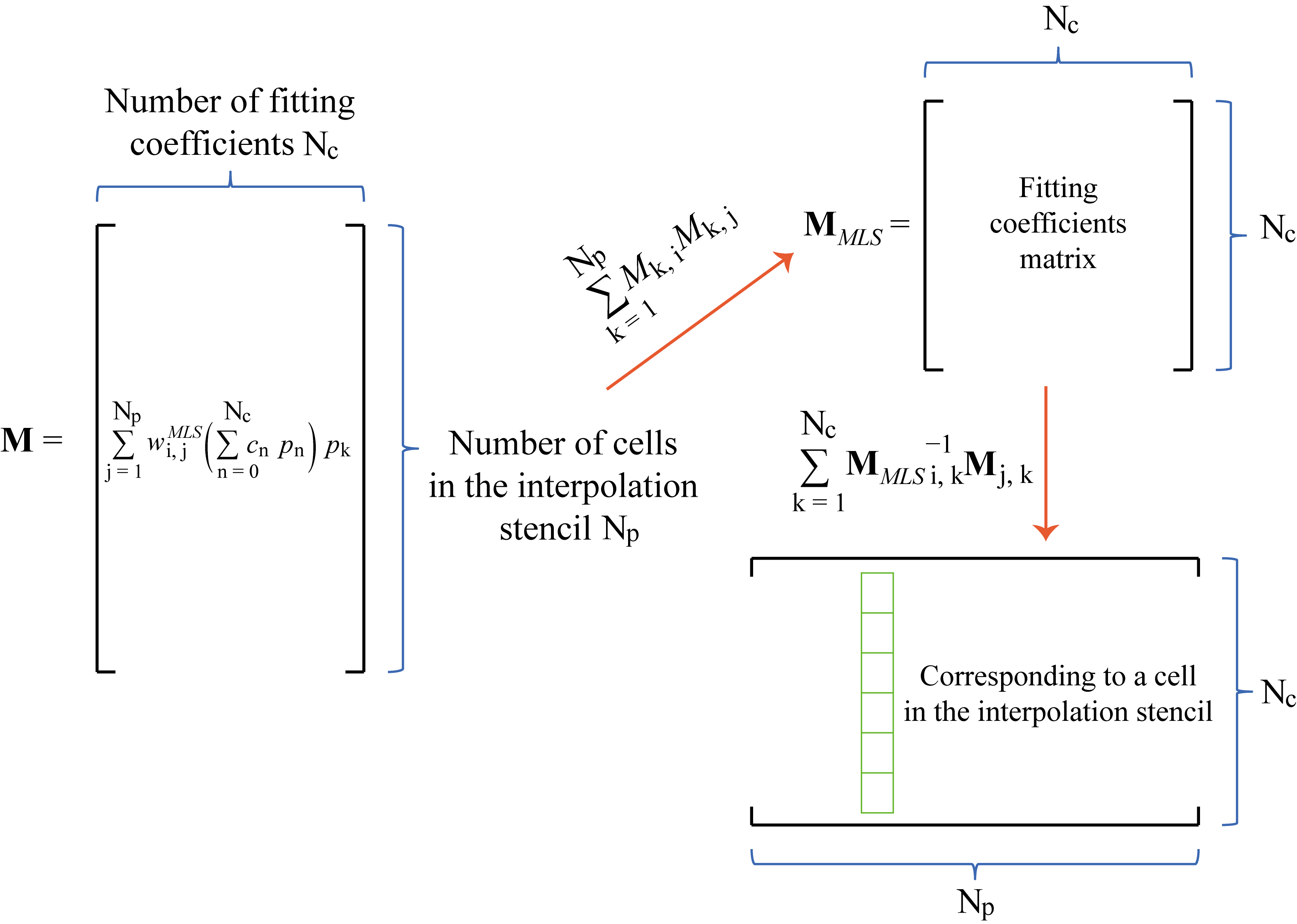}  
	\caption{Illustration of the matrix construction in the MLS method}
	\label{fig:MLS_matrix}
\end{figure}

\noindent Equation (\ref{eqn: MLS_error_matrix}) can be organised into a matrix form {\bf M}$_{MLS}\cdot {\bf c}={\bf b}$ (as illustrated in Fig. \ref{fig:MLS_matrix}), where {\bf M}$_{MLS}$ is the matrix of interpolation coefficients (a square matrix of order ${\rm N}_c$), ${\rm N_p}$ represents the total number of elements included in the interpolation template corresponding to the i$^{th}$ IB cell, and the source term {\bf b} is a vector with dimension of ${\rm N}_c$. Hence, {\bf c} can then be finally obtained by
\begin{equation} \label{eqn: MLS_c}
	{\bf c} = {\bf M}_{MLS}^{-1}\cdot{\bf b}.
\end{equation}

\noindent It should be noted that involving unknown variables in the right-hand side of Eq. (\ref{eqn: MLS_error_matrix}) may require iterations to obtain the optimised $c_{\rm k}$, thus the stencil in this work excludes other IB cells. The computation time for variable reconstruction can also be reduced.

\section*{Appendix B: Auxiliary modules for moving IB}\label{appB}
\setcounter{equation}{0}
\renewcommand{\theequation}{B.\arabic{equation}}
To facilitate the simulation of the free surface flow interacting with a moving structure (motion can either be prescribed or resolved in this work), a new dynamic mesh handling class accounted for moving IBs is required, along with the force calculation and motion state update modules. The forces acting on the structure {\bf F}$_{\scaleto{IB}{4.5pt}}$ and the resulting momentums {\bf T}$_{\scaleto{IB}{4.5pt}}$ are calculated with
\begin{eqnarray} \label{eqn: calcForce}
	{\bf F}_{\scaleto{IB}{4.5pt}}&=&\oiint{\bf f}_{\scaleto{IB}{4.5pt}}d\Gamma, \nonumber \\
	{\bf T}_{\scaleto{IB}{4.5pt}}&=&\oiint{\bf r}_{CoR}\times{\bf f}_{\scaleto{IB}{4.5pt}}d\Gamma.
\end{eqnarray}

\noindent In Eq. (\ref{eqn: calcForce}), {\bf f}$_{\scaleto{IB}{4.5pt}}$ is the force exerted on the inner IB faces and {\bf r}$_{CoR}$ is the vector pointing from the structure’s centre of rotation to the centre of the inner IB face. The motion state, i.e., the acceleration, velocity and the centre of rotation/mass of the structure is updated with these newly obtained {\bf F}$_{\scaleto{IB}{4.5pt}}$ and {\bf T}$_{\scaleto{IB}{4.5pt}}$ at the new time step via the Newmark scheme \cite{Newmark_1959}. 

After the IB moves to its new position, the updated boundary velocity of the structure at each triangular face centre is calculated as
\begin{equation} \label{eqn: movingIBVelocity}
	{\bf u}_{\Gamma, tri}^{\rm n+1} = {{\bf X}_{\Gamma, tri}^{\rm n+1}-{\bf X}_{\Gamma, tri}^{\rm n} \over {\rm\Delta} t}.
\end{equation}

\noindent where {\bf X}$_{\Gamma, tri}$ is the position vector of the STL triangular face centre. In order to obtain {\bf u}$_\Gamma^{\rm n+1}$, the weightings of velocity at these triangular faces’ centres associated with IB points need to be re-established in regard to distance-based weightings. The updated velocity {\bf u}$_\Gamma^{\rm n+1}$ is then passed to $c_0$ for the evaluation of {\bf u}$_{\scaleto{IB}{4.5pt}}$.

\end{document}